\documentclass[showpacs,preprintnumbers,amsmath,amssymb,prb,twocolumn,floatfix,amssymb,eqsecnum]{revtex4}
\usepackage[dvips]{graphicx,color}
\usepackage{dcolumn}
\usepackage{amsmath,amssymb,bbold,bm}

\begin{document}
\title{Weak Localization and Antilocalization in Topological Insulator Thin Films with Coherent Bulk-Surface Coupling} 
\author{Ion Garate and Leonid Glazman}
\affiliation{Department of Physics, Yale University, New Haven, CT 06520, USA}
\date{\today}
\begin{abstract}
We evaluate quantum corrections to conductivity in an electrically gated thin film of a three-dimensional (3D) topological insulator (TI).
We derive approximate analytical expressions for the low-field magnetoresistance as a function of bulk doping and bulk-surface tunneling rate. 
Our results reveal parameter regimes for both weak localization and weak antilocalization, and include diffusive Weyl semimetals as a special case.
\end{abstract}
\maketitle

\section{Introduction and Overview}

The theoretical discovery\cite{ti} of 3D topological insulators (TIs) in 2006 precipitated an avalanche of experiments aimed at detecting the signature behavior of these unconventional solids. 
Since then, angle-resolved photoemission spectra\cite{hasan2011} have given evidence for the Dirac-like dispersion and the momentum-dependent spin texture of TI surface states, whereas local STM probes have indicated a characteristic suppression of backscattering off surface imperfections.\cite{stm}
However, the most desired observation of a hallmark dc conduction confined to the surface layer of a 3D TI remains elusive.\cite{dimi2011}
The main problem is conduction through the bulk: 3D TIs are narrow-gap semiconductors, rich in bulk carriers that are either thermally activated and/or donated by crystalline lattice imperfections.
Along with attempts to reduce bulk charge carriers, experimentalists are developing techniques which allow to register a separate conduction channel on the surface of a 3D TI.\cite{kim2012}
Chief among these are measurements of low-field magnetoresistance combined with electrostatic gating of thin-film samples.\cite{wang2011,chen2010,checkelsky2011,he2011,chen2011,steinberg2011,hong2012}

Low-field magnetoresistance measurements unveil the interference correction $\delta\sigma$ to the Drude conductivity $\sigma_D$.\cite{inter}
At low temperatures, $\sigma_D$ is defined by independent acts of scattering of electrons off the crystal's imperfections, and is proportional to the classical diffusion constant $D$. 
When the phase relaxation length $l_\phi$ is parametrically longer than the scattering mean free path, quantum interference affects the conductivity to a measurable extent. 
The sign of the interference correction depends on the strength of spin-orbit interactions. 
For weak spin-orbit interactions ($l_{\rm so}\gg l_\phi$, where $l_{\rm so}$ is the spin-orbit scattering length), it follows that $\delta\sigma<0$. 
This is called weak localization (WL).
In contrast, strong spin-orbit interaction ($l_{\rm so}\ll l_\phi$) leads to suppression of backscattering and thus $\delta\sigma>0$. 
This is called weak antilocalization (WAL). 
Being interference effects, WL and WAL are degraded by a magnetic field $H$ when $H\gtrsim H_\phi\equiv \Phi_0/(8\pi l_\phi^2)$, where $\Phi_0=h/e$ is the flux quantum.
Yet, $\sigma_D$ is nearly immune to $H$ at such low fields.
Therefore, the low-field magnetoconductivity reads $\Delta\sigma(H)\equiv\sigma(H)-\sigma(0)\simeq\delta\sigma(H)-\delta\sigma(0)$.

All experiments to date report WAL in 3D TI thin films,\cite{kapitulnik2012} and ascribe it to the strong spin-orbit interaction in the electronic bands of these materials. 
For film thickness less than $l_\phi$, the measured $\Delta\sigma(H)$ agrees well with the functional form provided by 2D WAL theory, namely 
\begin{equation}
\label{eq:hikami}
\Delta\sigma(H)\simeq \alpha\,(e^2/2\pi^2 \hbar) f(H_\phi/H),
\end{equation}
 where $f(z)\equiv \ln z-\psi(1/2+z)$, with $\psi$ and $\alpha$ being the digamma function and a number,\cite{hikami1980} respectively.
In a system with a single conduction channel, $\alpha$ is universal and equals $1/2$.
The WAL contributions add for systems which are isolated from each other.
For example, having two independent parallel conduction channels yields $\alpha=1$, irrespective of the ratio of Drude conductivities of the two subsystems.

The relation between $\alpha$ and the number of parallel channels is at the heart of recent magnetoresistance experiments in 3D TIs.\cite{checkelsky2011,chen2011,steinberg2011}
Overall, the coefficient $\alpha$ is found to depend on the gate voltage.
For some devices,\cite{checkelsky2011,chen2011,steinberg2011} it changes from $\alpha=1/2$ all the way to $\alpha=1$.
A plausible interpretation for this variation is presented in Ref.~[\onlinecite{steinberg2011}].
At zero or positive bias applied to the top gate, electrons from the $n$-doped bulk reach the surface states easily: the entire film acts as a single electron system, and $\alpha=1/2$.
At negative bias, electrons are repeled from the top surface and, for strong enough bias, a depletion layer is formed adjacent to it. 
This depletion region separates the film into two subsystems: bulk carriers (combined with surface carriers from the bottom surface) on one side, and top-surface carriers on the other side. 
For a wide enough depletion layer, $\alpha=1$.

In spite of the ongoing scrutiny on the experimental front, quantum corrections to conductivity in 3D TIs have stimulated relatively little theoretical activity.
Even though the WAL contribution from TI surface states has been calculated explicitly,\cite{lu2011a,tkachov2011} there are no calculations that incorporate conducting 3D bulk states.
The main reason for this omission may be the prevailing view that quantum corrections originating from bulk TI states ought to be conceptually identical to those in ordinary strongly spin-orbit coupled systems, i.e. of WAL type.
Recently, an objection to this viewpoint has been raised,\cite{lu2011} declaring that quantum well states in ultrathin TI films may contribute via WL rather than WAL.
Although suggestive, the calculation of Ref.~[\onlinecite{lu2011}] is limited to quasi-2D films and disregards the coupling between bulk and surface states, which leaves out several experiments of interest.
Besides, its extrapolation to 3D bulk states has not been carried out properly.
 
In this paper we evaluate $\Delta\sigma$ for gated 3D thin films, as a function of the bulk carrier concentration and accounting for the coupling between surface and bulk states.
Our calculation applies to TI films that are thicker than the bulk mean free path, thinner than $l_\phi$, and not highly doped.
In these films, bulk carriers are three-dimensional and are concentrated around the $\Gamma$ point of the electronic band structure.
The resulting approximate analytical expressions for $\Delta\sigma$ (Eqs.~(\ref{eq:magres_bulk}),~(\ref{eq:res_tot}) and (\ref{eq:res_tot5})) are aimed at improving the interpretation of magnetoresistance measurements in TIs, in Weyl semimetals,\cite{burkov2011}  and in some class of topologically trivial materials. 
Although a few of our observations resemble those developed for graphene\cite{mccann2006} and 2D TIs,\cite{tkachov2011} there are qualitative differences originating from the 3D Dirac nature of bulk carriers in 3D TIs.

Altogether, the results reported here paint a richer picture than previously anticipated.
On one hand, we confirm the conventional crossover between $\alpha=1/2$ and $\alpha=1$ as a function of the gate voltage: the former corresponds to the case of coherently-coupled bulk and surface electron states, while the latter indicates a single decoupled Dirac cone on the top surface along with generic WAL from the rest of the film (containing coupled bulk and bottom surface).
On the other hand, less conventional results arise when the Fermi energy is close to the bulk band edge or when the Fermi energy is much larger than the bulk bandgap: in the former regime the bulk exhibits WL with $\alpha=-1$, whereas in the latter regime the bulk exhibits an anomalous WAL with $\alpha=1$. 
These two ``unusual'' bulk regimes, combined with the surface contributions, may result in a range of $\alpha$ including $\alpha<0$ and $\alpha>1$.

The rest of this work is organized as follows.
In Section II we evaluate quantum corrections to {\em bulk} conductivity. 
Readers not interested in technical details should read subsection IIA and quickly scan through IIB and IIC in order to get acquainted with the nomenclature; the main results of the section are collected in Section IID. 
The well-known message from IIA is that at low energies bulk electrons of TI films behave as massive 3D Dirac fermions with spin and valley (or orbital) degrees of freedom.
The direction of spin is locked with that of momentum, and valleys are coupled to one another by the mass of the Dirac fermions. 
The special case in which the Dirac mass vanishes is a time- and inversion-symmetric Weyl semimetal.

In Section IID we identify and count the number of ``soft'' Cooperon modes, which determine the magnitude and sign of $\Delta\sigma$ in the bulk.
Each soft Cooperon obeys a classical difussion equation and is thus associated with a conserved physical quantity.
Since charge is conserved, there is at least one soft Cooperon in (non-magnetic) bulk TIs. 
We find that additional soft Cooperons can emerge depending on the bulk doping concentration as well as the bulk bandgap. 
This realization leads to the most important results in IID, Eqs.~(\ref{eq:res_bulk})-(\ref{eq:magres_bulk}), which indicate that for bulk states $\alpha$ may acquire three different universal values.
On one hand, WL with $\alpha=-1$ is possible when the bulk Fermi surface is ``small'' (as defined in the text), because in this case the spin-momentum locking of bulk states becomes weak and the spin of electrons is nearly conserved. 
In contrast, WAL with $\alpha=1$ can arise for bulk TIs with particularly small bandgaps, because in such case bulk electrons can be described by a 3D analogue of graphene with two nearly decoupled valleys, each contributing $1/2$ to $\alpha$.
For a more generic case, in which neither valley nor spin are approximately conserved, the quantum interference is similar to that of an ordinary film with strong spin-orbit coupling and therefore $\alpha=1/2$.  
Magnetic fields perpendicular to the TI film can be used to induce crossovers between different universal regimes  of $\alpha$.
The accessible values of $\alpha$ and the corresponding crossover fields depend on the bulk electron density.

In Section III we evaluate the {\em full} $\Delta\sigma$ in 3D TI thin films, which comprises coupled bulk and surface contributions.
Sections IIIA and IIIB cover preliminary material that is needed to derive the main results in IIIC.
Section IIIA reviews the well-established fact that, in absence of magnetic order, isolated TI surface states exhibit WAL with $\alpha=1/2$ (in this paper we assume one Dirac cone per surface).
Section IIIB develops a diagrammatic framework for evaluating quantum corrections to conductivity in ordinary tunnel-coupled layers. 
Readers who are not interested in technicalities can disregard the diagrams in the figures and concentrate on the outcome of the calculation (Eqs.~(\ref{eq:q1})-(\ref{eq:Dii})), as well as on the subsequent discussion.
One qualitative point made therein is that the crossover from weak to strong coupling (which is accompanied by a change in $\alpha$ from $1$ to $1/2$)  occurs when the interlayer resistance for a square of area $l_\phi^2$ becomes smaller than the sum of the classical intralayer resistances.
	
Section IIIC combines results from IID, IIIA and IIIB in order to figure out quantum corrections to conductivity in experimentally realized TI films.
The most important results in IIIC are Eqs.~(\ref{eq:res_tot}) and ~(\ref{eq:res_tot5}), which describe how $\Delta\sigma$ depends on the bulk doping concentration, on the phase relaxation rate, and on the bulk-surface tunneling rate.
Some special cases of these results are highlighted in Appendix~\ref{sec:special}.
A salient conclusion is that the WL regime of isolated bulk states is generally eliminated when either one of the film surfaces is strongly coupled to bulk states, in which case the film displays $1/2\leq\alpha\leq 1$. 
However, WL can still be present if the TI surfaces have short phase relaxation lengths. 
  

Finally, Section IIID characterizes the electrostatics of the depletion layer and estimates the bulk-surface tunneling rate in TI films.
This estimate confirms experimental indications showing that both weak and strong bulk-surface coupling are accessible by mediation of a gate voltage.

\section{Quantum Corrections to Bulk Conductivity}

This section is devoted to evaluating $\delta\sigma$ for the bulk states of a 3D TI. 
As a byproduct, we derive $\delta\sigma$ for a time-reversal symmetric Weyl semimetal.
The contribution from TI surface states will be discarded until the next section.

\subsection{Model}

The bulk band structure of a 3D TI near the $\Gamma$ point can be approximated by the following ${\bf k}\cdot{\bf p}$ Hamiltonian:~\cite{zhang2009}
\begin{align}
\label{eq:model_b}
&{\cal H}=\sum_{\bf k}\Psi^\dagger_{\bf k} h({\bf k}) \Psi_{\bf k}\nonumber\\
& h({\bf k}) \simeq \epsilon({\bf k}){\bf 1}_4+M({\bf k}){\bf 1}_2\,\tau^z+\hbar\left(v_z k_z\sigma^z+v_\perp {\bf k}_\perp\cdot{\boldsymbol \sigma}^\perp\right)\tau^x, 
\end{align}
where ${\boldsymbol \tau}$ is an orbital pseudospin ($\tau^z=T,B$), ${\boldsymbol \sigma}$ is the real spin ($\sigma^z=\uparrow,\downarrow$), ${\bf k}=({\bf k}_\perp,k_z)$ is the momentum measured from the $\Gamma$ point of the Brillouin zone, ${\bf 1}_N$ is an $N\times N$ identity matrix, $\Psi=(\Psi_{T\uparrow},\Psi_{T\downarrow},\Psi_{B\uparrow},\Psi_{B\downarrow})$ is a 4-spinor,
$\epsilon({\bf k})=\epsilon(-{\bf k})$ is the part of the Hamiltonian that is independent of spin/pseudospin indices, $v_z$ and $v_\perp$ are the Fermi velocities, and $M({\bf k})=M_0-M_1 k_\perp^2-M_2 k_z^2$ is the mass term (independent of spin). 
$M_0$, $M_1$ and $M_2$ are constants.

Equation~(\ref{eq:model_b}) captures the bottom of the conduction band and the top of the valence band in the vicinity of the $\Gamma$ point ($k\equiv 0$), where the bandgap is smallest. 
It models 3D Dirac fermions with a Dirac mass that equals half the energy gap. 
For the purposes of this paper we ignore $\epsilon({\bf k})$, and assume $M({\bf k})=M={\rm const}>0$ as well as spherical symmetry ($v_z=v_\perp=v$). 
These assumptions simplify calculations without incurring in qualitative loss of generality. 
For instance, the XXZ anisotropy can be modeled by promoting the diffusion constant from a scalar to a matrix. 
Also, the $k^2$ terms in $M({\bf k})$ can be incorporated into our final results by $M\to |M({\bf k}_F)|$, where ${\bf k}_F$ is the Fermi wave vector.
Note that in absence of spherical symmetry the Fermi surface does not have a constant mass; this complication will be disregarded in the present paper. 
Finally, $\epsilon({\bf k})$ can be absorbed into the definition of the Fermi energy. 

\begin{figure}
\begin{center}
\includegraphics[scale=0.35]{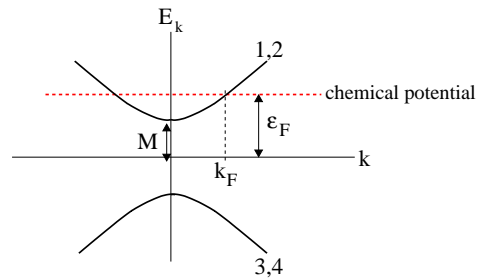}
\caption{Bulk energy bands of an $n$-doped 3D TI near the $\Gamma$ point, in the spherical approximation. The momentum $k$ is measured from the $\Gamma$ point. The energies $\epsilon_F$ and $M$ are measured with respect to midgap.}
\label{fig:bands}
\end{center}
\end{figure}

The energy eigenvalues for $h({\bf k})$ in the spherical approximation are $E_{{\bf k}\pm}=\pm\sqrt{\hbar^2v^2 k^2+M^2}$, each doubly degenerate (Fig.~\ref{fig:bands}).
The corresponding Bloch states can be written as 
\begin{equation}
\label{eq:eigen0}
|\Psi_{{\bf k}\alpha}\rangle=(1/\sqrt{V})\exp(i {\bf k}\cdot{\bf r})|\alpha {\bf k}\rangle, 
\end{equation}
where $V$ is the volume of the TI and $\alpha\in\{1, 2, 3, 4\}$ is a band index ($1$ and $2$ denote conduction bands, while  $3$ and $4$ denote valence bands).
This $\alpha$ is obviously unrelated to that of Eq.~(\ref{eq:hikami}); from here on it will be clear from the context which one we are referring to.
For concreteness we set the chemical potential in the bulk conduction band, although all results obtained below will be directly applicable to $p$-doped bulk TIs as well.
The density of conduction band electrons is then
\begin{equation}
\label{eq:n}
n\simeq \frac{\left(\epsilon_F^2-M^2\right)^{3/2}}{\pi^2\hbar^3 v^3},
\end{equation}
where $\epsilon_F$ is the Fermi energy measured from the middle of the bulk energy gap. 
Adopting the basis $\{|T\uparrow\rangle, |T\downarrow\rangle,|B\uparrow\rangle,|B\downarrow\rangle\}$, the two eigenspinors corresponding to the conduction bands near the $\Gamma$ point are
\begin{eqnarray}
\label{eq:eigenstates}
|1 {\bf k}\rangle&=&\sqrt{\frac{E_k+M}{2 E_k}}\left(1,0,\frac{\hbar v k_z}{E_k+M},\frac{\hbar v k_+}{E_k+M}\right)\nonumber\\
|2 {\bf k}\rangle&=&\sqrt{\frac{E_k+M}{2 E_k}}\left(0,1,\frac{\hbar v k_-}{E_k+M},\frac{-\hbar v k_z}{E_k+M}\right),
\end{eqnarray} 
where $k_\pm=k_x\pm i k_y$ and $E_k=E_{{\bf k},+}$.
Since all non-Hall dc transport properties of good conductors involve states close to the Fermi energy, we hereafter ignore valence bands.

Unlike in the ${\bf k}\cdot{\bf p}$ Hamiltonians for graphene and 2D (or quasi-2D) TIs, Eq.~(\ref{eq:model_b}) cannot be decomposed into two $2\times 2$ block diagonal matrices (due to $M\neq 0$).
In addition, the $k_z$ band dispersion absent in 2D cannot be neglected in our case. 
These two features make the calculations and results of this section quite different from those of Refs.~[\onlinecite{mccann2006,tkachov2011,lu2011}].

Equation~(\ref{eq:model_b}) becomes inaccurate when the chemical potential moves up in the conduction band and electron pockets away from the $\Gamma$ point begin to be populated.
These additional pockets contribute to quantum interference, and the total $\delta\sigma$ depends on the scattering rate between different electron pockets.
Although a realistic study of the full band structure is beyond the scope of this paper, we expect calculations based on Eq.~(\ref{eq:model_b}) to provide a generic understanding of quantum corrections to conductivity in 3D Dirac materials at low-to-moderate doping concentrations.

\subsection{Formalism}

In order to quantify the conductivity of a bulk TI, we begin by characterizing the simplest possible disorder potential: $V_{\rm dis}({\bf r})=V({\bf r}){\bf 1}_4$, which is time-independent (elastic) and independent of spin as well as orbital degrees of freedom. 
For simplicity we assume $V({\bf r})$ to be slowly-varying at the atomic scale, yet short-ranged compared to the mean free path: $V({\bf r})=V_0\delta({\bf r})$. 
It is due to its slow spatial variation on atomic lenghtscales that $V_{\rm dis}$ becomes an identity operator in orbital space.
With such disorder realization, the Fermi-surface lifetime $\tau_0$ for the $\alpha=1,2$ eigenstates in Eq.~(\ref{eq:eigen0}) obeys
\begin{align}
\label{eq:lifetime}
\frac{1}{\tau_0} &=\frac{2\pi u_0}{\hbar}\int_{{\bf k}'}\sum_{\alpha'}|\langle\alpha {\bf k}_F|\alpha' {\bf k}_F'\rangle|^2 \delta(\epsilon_F-E_{{\bf k}'\alpha'})\nonumber\\
&\simeq\frac{\pi u_0 \nu}{\hbar} \left(1+\frac{M^2}{\epsilon_F^2}\right),
\end{align}
where $\int_{\bf k}\equiv\int d^3 k/(2\pi)^3$,  $u_0\equiv n_i V_0^2$, $n_i$ is the density of impurities,
and $\nu$ is the density of states per band and per unit volume at $\epsilon_F$.

A related quantity is the transport scattering rate $\tau^{-1}$, 
\begin{align}
\frac{1}{\tau} &\equiv \frac{2\pi u_0}{\hbar}\int_{{\bf k}'}\sum_{\alpha'} (1-\hat{{\bf k}}_F\cdot\hat{{\bf k}}_F') |\langle\alpha {\bf k}_F|\alpha' {\bf k}_F'\rangle|^2 \delta(\epsilon_F-E_{{\bf k}'\alpha'})\nonumber\\
&= \frac{2}{3\tau_0}\frac{\epsilon_F^2+2 M^2}{\epsilon_F^2+M^2}.
\end{align}
The momentum-dependence of $|\alpha {\bf k}\rangle$ makes $\tau_0\neq\tau$ even for $\delta$-function impurity potentials.
Throughout this work we impose $(\epsilon_F-M)\tau\gg \hbar$ or equivalently $k_F l\gg 1$, where $l=(D\tau)^{1/2}$ is the elastic mean free path, 
\begin{equation}
k_F=(\epsilon_F^2-M^2)^{1/2}/(\hbar v)
\end{equation}
is the Fermi wave vector and
\begin{equation}
D=v_F^2\tau/3=v^2\tau(1-M^2/\epsilon_F^2)/3
\end{equation}
is the classical diffusion constant.
 
Next, we consider a TI with spatial dimensions $L\times L$ in the $xy$ plane and a thickness $W$ along the $z$ direction. 
We take a thin film geometry with $L\gg l_\phi\gg l$ and $l_\phi\gg W\gg l$, where $l_\phi=(D\tau_\phi)^{1/2}$ is the coherence length and $\tau_\phi$ is the phase relaxation time. 
The conductivity of this film is 
\begin{equation}
\sigma=\sigma_D+\delta\sigma,
\end{equation}
where $\sigma_D$ is the classical (Drude) part and $\delta\sigma$ is the part coming from quantum interference.

On one hand, the Drude conductivity can be approximated as
\begin{equation}
\label{eq:sd}
\sigma_D\simeq\frac{e^2\hbar}{2\pi}\sum_{\alpha,\beta}\int_{\bf k} v^x_{\alpha\beta}({\bf k})\tilde{v}^x_{\beta\alpha}({\bf k}) G^R_\alpha({\bf k}) G^A_\beta({\bf k}),
\end{equation}
where we have assumed a spatially uniform dc electric field and $\alpha,\beta\in\{1,2\}$.
$v^x_{\alpha\beta}=\langle\alpha{\bf k}|{\bf v}\cdot\hat{x}|\beta {\bf k}\rangle$ is a matrix element for the $x$-component of the bare velocity operator ${\bf v}=v\tau^x{\boldsymbol\sigma}$, which obeys 
\begin{equation}
\label{eq:vbare}
{\bf v}_{\alpha\beta}({\bf k})=\hbar v^2 ({\bf k}/E_k) \delta_{\alpha\beta}\,\,\,\mbox{(for $\alpha,\beta\in\{1,2\}$)}. 
\end{equation}
Disorder vertex corrections renormalize Eq.~(\ref{eq:vbare}) to  
\begin{equation}
\tilde{{\bf v}}_{\alpha\beta}={\bf v}_{\alpha\beta}(\tau/\tau_0),
\end{equation}
see Appendix~\ref{sec:ren}.
In addition,
\begin{equation}
\label{eq:ds}
G_\alpha^{R(A)}({\bf k})= \left[\epsilon_F-E_{{\bf k}\alpha}+(-) \frac{i\hbar}{2\tau_0}\right]^{-1}
\end{equation}
is the zero-frequency retarded (advanced) Green's function in the band eigenstate basis.
Using $G^{R(A)}_1({\bf k})=G^{R(A)}_2({\bf k})\equiv G^{R(A)}({\bf k})$, Eq.~(\ref{eq:sd}) yields 
\begin{equation}
\sigma_D=2 e^2 \nu D.
\end{equation}

\begin{figure}
\begin{center}
\includegraphics[scale=0.35]{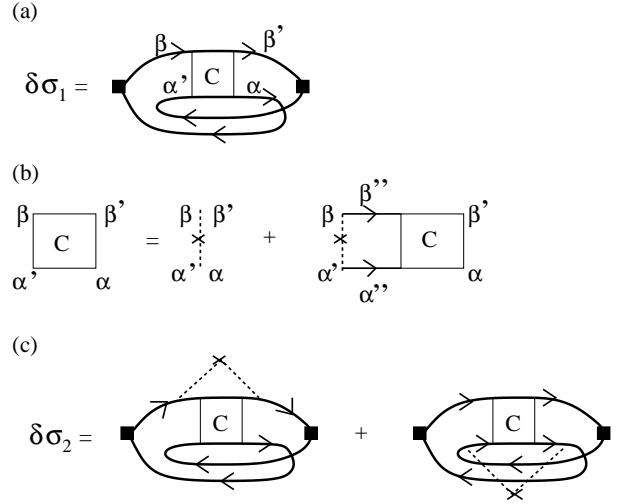}
\caption{(a) Feynman diagram for $\delta\sigma_1$, defined in the text. Filled squares denote velocity operators (including disorder vertex corrections), $C$ is the Cooperon. (b) Diagrammatic representation of the Bethe-Salpeter equation for the Cooperon. Crosses correspond to impurity scattering centers. Solid lines with arrows are disorder-averaged Green's functions. (c) Additional Feynman diagrams that contribute to conductivity of 3D TIs even when impurity scattering is isotropic.}
\label{fig:cofig}
\end{center}
\end{figure}

On the other hand, the quantum correction $\delta\sigma$ can be written as $\delta\sigma\simeq\delta\sigma_1+\delta\sigma_2$, represented pictorially in Fig.~\ref{fig:cofig}.
Following standard approximations, the expression for $\delta\sigma_1$ is
\begin{align}
\label{eq:ds}
\delta\sigma_1 &\simeq \frac{e^2\hbar}{2\pi}\sum_{\alpha,\alpha',\beta,\beta'}\int_{\bf k} \tilde{v}^x_{\alpha\beta}({\bf k}) \tilde{v}^x_{\beta'\alpha'}(-{\bf k}) G^R_\alpha({\bf k}) G^R_{\alpha'}(-{\bf k})\nonumber\\
&~~~\times G^A_\beta({\bf k}) G^A_{\beta'}(-{\bf k})\frac{1}{W}\int \frac{d^2 Q}{(2\pi)^2} C^{\beta\beta'}_{\alpha'\alpha}({\bf k},{\bf k},{\bf Q}).
\end{align}
In the second line of Eq.~(\ref{eq:ds}) we have exploited the condition $W\ll l_\phi$, which allows to set $Q_z=0$ everywhere.
$C^{\beta\beta'}_{\alpha'\alpha}({\bf k}_1,{\bf k}_2,{\bf Q})$ are the matrix elements of the Cooperon matrix $\hat{C}$ in the band eigenstate basis. 
${\bf Q}=(Q_x,Q_y)$ is the momentum of the Cooperon, whose magnitude ranges from $0$ to $\simeq (D\tau)^{-1/2}$.
$\hat{C}$ obbeys the Bethe-Salpeter equation (Fig.~\ref{fig:cofig}b):
\begin{align}
\label{eq:bs}
&C^{\beta\beta'}_{\alpha'\alpha}({\bf k}_1,{\bf k}_2,{\bf Q})=\Gamma^{\beta\beta'}_{\alpha'\alpha}({\bf k}_1,{\bf k}_2,{\bf Q})+ \int_{{\bf k}_3}\Gamma^{\beta\beta''}_{\alpha'\alpha''}({\bf k}_1,{\bf k}_3,{\bf Q})\nonumber\\
&~~~~~~~\times G^A_{\beta''}({\bf k}_3) G^R_{\alpha''}(-{\bf k}_3+{\bf Q}) C^{\beta''\beta'}_{\alpha''\alpha}({\bf k}_3,{\bf k}_2,{\bf Q}),
\end{align}
where a sum over repeated indices is implied and
\begin{equation}
\Gamma^{\beta\beta'}_{\alpha'\alpha}({\bf k}_1,{\bf k}_2,{\bf Q})\equiv u_0\langle\beta {\bf k}_1|\beta' -{\bf k}_2+{\bf Q}\rangle\langle\alpha' -{\bf k}_1+{\bf Q}|\alpha {\bf k}_2\rangle\nonumber
\end{equation}
is the bare disorder vertex (first term on the right hand side of Fig.~\ref{fig:cofig}b).

Equation~(\ref{eq:bs}) is a complicated integral equation because $C^{\beta\beta'}_{\alpha'\alpha}$ is a function of three momenta.
This is unlike in simplest examples, where the Cooperon depends only on ${\bf Q}$.
The difficulty originates from the momentum-dependence of $|\alpha {\bf k}\rangle$, which cannot be overlooked as it crucially determines both the magnitude and the sign of $\delta\sigma$.
One procedure\cite{garate2009} to solve Eq.~(\ref{eq:bs}) starts by writing the Cooperon in the two-particle spin/orbit basis $\{|m,m'\rangle\}$, where $m\in\{T\uparrow,T\downarrow,B\uparrow,B\downarrow\}$:
\begin{align}
\label{eq:trans}
&C^{\beta \beta'}_{\alpha' \alpha}({\bf k}_1, {\bf k}_2, {\bf Q})=\sum_{m,m',n,n'}\langle\alpha',-{\bf k}_1+{\bf Q}|m'\rangle\langle\beta {\bf k}_1|m\rangle\nonumber\\
&\times\langle n|\beta',-{\bf k}_2+{\bf Q}\rangle\langle n'|\alpha {\bf k}_2\rangle C^{m n }_{m' n'}({\bf Q}).
\end{align}
We then make the ansatz that $C^{m n}_{m' n'}$ depends on ${\bf Q}$ but not on ${\bf k}_1$ and ${\bf k_2}$; the entire ${\bf k}_1$- and ${\bf k}_2$-dependence of $C^{\beta \beta'}_{\alpha' \alpha}({\bf k}_1,{\bf k}_2,{\bf Q})$ originates from the overlap matrix elements of Eq.~(\ref{eq:trans}).
The internal consistency of this ansatz can be demonstrated by substituting Eq.~(\ref{eq:trans}) in Eq.~(\ref{eq:bs}), which produces an algebraic equation for $C^{m n}_{m' n'}$ that is more tractable than the original integral equation:
\begin{equation}
\label{eq:C_mat}
C^{m n}_{m' n'}({\bf Q})= u_0 \delta_{m n}\delta_{m' n'}+\sum_{l,l'}U^{m l}_{m' l'}({\bf Q}) C^{l n}_{l' n'}({\bf Q}),
\end{equation}
where
\begin{equation}
\label{eq:U}
U^{m l}_{m' l'}({\bf Q})=u_0\int \frac{d^3 k}{(2\pi)^3} G_{m l}^A({\bf k})G^R_{m' l'}(-{\bf k}+{\bf Q})
\end{equation}
and 
\begin{equation}
G^{R (A)}_{m l}({\bf k})=\sum_\alpha \langle m|\alpha {\bf k}\rangle G_\alpha^{R (A)}({\bf k}) \langle\alpha {\bf k}|l\rangle.
\end{equation}
In matrix language, Eq.~(\ref{eq:C_mat}) can be rewritten as
\begin{equation}
\label{eq:C_mat2}
\hat{C}=({\bf 1}_{16}-\hat{U})^{-1} \hat{C}^{(0)},
\end{equation}
where $\hat{C}^{(0)}=u_0 {\bf 1}_{16}$.
Once we obtain $C^{m n}_{m' n'}$, we use Eq.~(\ref{eq:trans}) in order to recover $C^{\beta\beta'}_{\alpha'\alpha}$. 
During this operation we neglect ${\bf Q}$ in the overlap matrix elements, 
which is a good approximation because $\delta\sigma$ is dominated by elements of $C^{m n}_{m' n'}({\bf Q})$ that are strongly peaked at $Q\simeq 0$.

For $\epsilon_F$ in the conduction band, we once again limit ourselves to $\alpha,\beta,\alpha',\beta'\in\{1,2\}$ in Eq.~(\ref{eq:ds}).
Then we can approximate ${\bf k}\simeq {\bf k}_F$ inside the Cooperon, and an integration over $|{\bf k}|$ yields  
\begin{equation}
\label{eq:ds2bis}
\delta\sigma_1 \simeq  -6 \frac{e^2}{\hbar^2}  \nu D \tau\tau_0\frac{1}{W}\int\frac{d^2 Q}{(2\pi)^2}\overline{C}({\bf Q}),  
\end{equation}
where
\begin{equation}
\label{eq:overline_c}
\overline{C}({\bf Q}) \equiv \int \frac{d\Omega_{\bf k}}{4\pi}\hat{\bf k}_x^2\sum_{\alpha,\alpha'=1,2}C^{\alpha \alpha'}_{\alpha' \alpha}({\bf k}_F,{\bf k}_F,{\bf Q})
\end{equation}
and $d\Omega_{\bf k}$ is the differential solid angle subtended by $\hat{{\bf k}}$.

Note that $\delta\sigma_1$ depends on the lifetime $\tau_0$ of Bloch states as well as on the transport relaxation time $\tau$. 
As mentioned above, the difference between $\tau$ and $\tau_0$ comes from the momentum dependence of $|\alpha {\bf k}\rangle$ states. 
At any rate, the full correction $\delta\sigma$ depends only on $\tau$ due to the additional contribution from $\delta\sigma_2$ (see Fig.~\ref{fig:cofig}c and Eq.~(\ref{eq:nasty})). 
Equation~(\ref{eq:nasty}) can be evaluated using the same procedure as for $\delta\sigma_1$.
For instance, in Appendix~\ref{sec:ds2} we derive 
\begin{equation}
\label{eq:ds2}
\delta\sigma_2\simeq\left\{\begin{array}{ccc} 0 & {\rm if } & (\epsilon_F-M)/M\ll 1\\
				 -(1/3)\delta\sigma_1 & {\rm if }& (\epsilon_F-M)/M\gg 1.\\   
\end{array}\right.
\end{equation}        
The full form of the quantum correction, $\delta\sigma_1+\delta\sigma_2$, depends only on the transport mean free path $\tau$ and has (in appropriate limits) a universal magnitude, see Eqs.~(\ref{eq:res_bulk}) and ~(\ref{eq:magres_bulk}).

\subsection{Calculations}

The road map to $\delta\sigma$ starts from a calculation of $\hat{U}$ in Eq.~(\ref{eq:U}).
In Appendix \ref{sec:u} we derive
\begin{align}
\label{eq:coeffs}
U^{m l}_{m' l'} &= a\,\delta_{m l}\delta_{m' l'}+ \sum_\mu b_\mu\,\Lambda^\mu_{m' l'}\delta_{m l}\nonumber\\
&+\sum_\mu c_\mu\,\Lambda^\mu_{m l}\delta_{m' l'}+\sum_{\mu,\nu} d_{\mu\nu}\,\Lambda^\mu_{m l}\Lambda^\nu_{m' l'},
\end{align}
where $\mu,\nu\in\{1,2,3,4\}$, $\Lambda^i=\sigma^i\tau^x$ for $i\in\{1,2,3\}$ and $\Lambda^4={\bf 1}_{2}\,\tau^z$.
In addition, $a$, $b_\mu$, $c_\nu$ and $d_{\mu\nu}$ are ${\bf Q}$-dependent coefficients whose explicit expressions are shown in Appendix~\ref{sec:u}.
With those, $\hat{U}$ is fully determined.

The next task is to get $C^{m n}_{m' n'}({\bf Q})$ from Eq.~(\ref{eq:C_mat2}).
While $({\bf 1}_{16}-\hat{U}({\bf Q}))$ can be inverted numerically, it is not possible to do so analytically for $Q\neq 0$.
Since we are interested in analytical expressions, we follow an approximate three-step route.

First, we diagonalize $({\bf 1}_{16}-\hat{U})$ for $Q=0$, analytically.
All eigenvalues can be written in the form $\Delta_g\tau_0$, 
where $\Delta_g$ is the ``intrinsic'' Cooperon gap or mass.
We find that one of the eigenvalues has $\Delta_g=0$ for any $\epsilon_F$ and $M$, which is a reflection of combined time-reversal symmetry and charge conservation.
As we elaborate in the next subsection, there may be additional eigenvalues with $\Delta_g\simeq 0$ when $(\epsilon_F-M)/M\ll 1$ and $(\epsilon_F-M)/M\gg 1$. 
Hereafter we refer to eigenvectors of $\Delta_g\simeq 0$ eigenvalues as gapless (or massless, or ``soft'') modes.
Because $\Delta_g\simeq 0$ eigenvalues make $\hat{C}$ large, $\delta\sigma$ is determined mainly by soft modes.

Second, we extrapolate the $Q=0$ case to $Q\neq 0$ perturbatively, with the objective of finding how the eigenvalues of the gapless modes depend on $Q$. 
To that end $\delta \hat{U} ({\bf Q})\equiv \hat{U}({\bf 0})-\hat{U}({\bf Q})$ is written in the basis that diagonalizes $\hat{U}({\bf 0})$. 
The shift of $Q=0$ eigenvalues under $\delta\hat{U}({\bf Q})$ is then evaluated through standard second order perturbation theory.
The need to go to second order in $\delta \hat{U}$ originates from the fact that several matrix elements of $U^{m n}_{m' n'}({\bf Q})$ are linear in $Q$ (see Appendix~\ref{sec:u}). 
When $(\epsilon_F-M)/M\ll 1$ and $(\epsilon_F-M)/M\gg 1$, perturbation theory leads to eigenvalues $(D Q^2+\Delta_g)\tau_0$.
The fact that $D$ contains the transport time $\tau$ rather than the scattering time $\tau_0$ is generally crucial in order to arrive at the correct result for $\delta\sigma$.

Third, we invert the diagonalized matrix, and transform its outcome to the $|m,m'\rangle$ basis by using the $Q=0$ eigenvector matrix (the change of unperturbed eigenvectors under $\delta\hat{U}({\bf Q})$ is deemed unimportant.)
This yields $C^{m n}_{m' n'}({\bf Q})$.

Once we have $C^{m n}_{m' n'}({\bf Q})$, we use Eq.~(\ref{eq:trans}) in order to extract $C^{\alpha\beta}_{\beta\alpha}({\bf k},{\bf k}',{\bf Q})$.
This is then plugged in Eqs.~(\ref{eq:ds2bis}) and ~(\ref{eq:nasty}).

\subsection{Results}

The diagonalization of Eq.~(\ref{eq:C_mat2}) at $Q=0$ shows one genuinely gapless Cooperon mode ($\Delta_g=0$, c.f. Sec. IIC), with a spin-singlet and orbital-triplet eigenvector:
\begin{equation}
\label{eq:g1}
\left[\frac{\epsilon_F+M}{2\sqrt{\epsilon_F^2+M^2}}|T T\rangle+ \frac{\epsilon_F-M}{2\sqrt{\epsilon_F^2+M^2}}|B B\rangle\right]\left(|\uparrow\downarrow\rangle-|\downarrow\uparrow\rangle\right).
\end{equation}
The fact that Eq.~(\ref{eq:g1}) remains gapless for any $\epsilon_F/M$ is a physical manifestation of charge conservation.
This situation differs qualitatively from 2D TIs in HgTe quantum wells,\cite{tkachov2011} where a nonzero mass term gaps all Cooperons. 
The reason for the difference is that in 2D TIs the mass term acts somewhat like a Zeeman field in a 2D electron gas with Rashba spin-orbit interaction.

Importantly, the diagonalization of Eq.~(\ref{eq:C_mat2}) reveals two qualitatively distinct regimes of quantum interference, $(\epsilon_F-M)/M\ll 1$ and $(\epsilon_F-M)/M\gg 1$, which potentially host additional gapless Cooperon modes.
As we discuss below, these additional gapless modes can change and even reverse the contribution to $\delta\sigma$ coming from Eq.~(\ref{eq:g1}).

When $(\epsilon_F-M)/M\gg 1$, we identify a slightly gapped (soft) Cooperon mode with 
\begin{equation}
\label{eq:tau_s}
\Delta_g=2 (M^2/\epsilon_F^2)\tau_0^{-1}\equiv\tau_v^{-1}\ll\tau_0^{-1},
\end{equation}
 whose eigenvector is a spin-singlet and an orbital-triplet:
\begin{equation}
\label{eq:g2}
\frac{1}{2}\left(|T B\rangle+|B T\rangle\right)\left(|\uparrow\downarrow\rangle-|\downarrow\uparrow\rangle\right).
\end{equation}
Physically, $\tau_v^{-1}$ is the rate of ``intervalley'' transitions ($|T\rangle+|B\rangle \to |T\rangle-|B\rangle$) induced by the ``mass term'' ($M\tau^z$) in Eq.~(\ref{eq:model_b}). 
Because both Eq.~(\ref{eq:g1}) and Eq.~(\ref{eq:g2}) are spin-singlets, their contributions to $\delta\sigma$ are of WAL type (this is proven below).

Incidentally, $M=0$ is the physically relevant regime for Weyl semimetals, which have two degenerate Dirac points with linear energy dispersion along the three momenta axes.
Unlike in graphene,\cite{mccann2006} where there are $4$ gapless Cooperon modes (in absence of atomically sharp defects and hexagonal warping), in a Weyl semimetal we obtain only $2$ gapless Cooperon modes.
This difference stems from the fact that the SU(2) ``valley symmetry'' of graphene\cite{mccann2006} gets reduced to a U(1) symmetry in Weyl semimetals, due to the band dispersion along $z$. 
Acting somewhat like a Zeeman field would in a free electron gas, the $k_z$ dispersion generates a mass for orbital-singlet modes, which is why the nearly-gapless Cooperons in Eq.~(\ref{eq:g1}) and ~(\ref{eq:g2}) are orbital-triplets.

When $(\epsilon_F-M)/M\ll 1$, there are three soft modes with gap 
\begin{equation}
\label{eq:tau_v}
\Delta_g=(2/9)(1-M/\epsilon_F)^2\tau_0^{-1}\equiv\tau_s^{-1}\ll\tau_0^{-1}.
\end{equation} 
Physically,  $\tau_s^{-1}$ is the rate of spin-flip transitions induced by the ``spin-orbit term'' ($v {\bf k}\cdot{\boldsymbol\sigma} \tau^x$) in Eq.~(\ref{eq:model_b}). 
The eigenvectors for the three slightly gapped modes are
\begin{align}
\label{eq:trip}
&(\lambda_1|T T\rangle+\lambda_2|B B\rangle)|\downarrow\downarrow\rangle\nonumber\\
& (\lambda_1|T T\rangle+\lambda_2|B B\rangle)|\uparrow\uparrow\rangle\nonumber\\
& (\lambda_3|T T\rangle +\lambda_4|B B\rangle)\left(|\uparrow\downarrow\rangle+\downarrow\uparrow\rangle\right),
\end{align}
where $\lambda_1,...,\lambda_4$ are coefficients that depend only on $\epsilon_F/M$, such that $\lambda_1\simeq\lambda_3\simeq 1+O[(\epsilon_F/M-1)^2]$ and $\lambda_2\simeq\lambda_4\simeq O[(\epsilon_F/M-1)]$. 
Therefore, the three soft modes in Eq.~(\ref{eq:trip}) are all spin and orbital triplets.
As will be demonstrated momentarily, their contribution to $\delta\sigma$ is of WL type.

Next we determine $\overline{C}$ (c.f. Eq.~(\ref{eq:overline_c})) by diagonalizing Eq.~(\ref{eq:C_mat}) at $Q\neq 0$ and doing the angular integration in Eq.~(\ref{eq:overline_c}).
For $(\epsilon_F-M)/M\ll 1$ we obtain
\begin{equation}
\label{eq:cav1}
\overline{C}\simeq  \frac{\hbar}{6\pi\nu\tau^2}\left[-\frac{1}{D Q^2+\tau_\phi^{-1}}+\frac{3}{D Q^2+\tau_\phi^{-1}+\tau_s^{-1}}\right].
\end{equation}
For $(\epsilon_F-M)/M\gg 1$, we instead get
\begin{equation}
\label{eq:cav2}
\overline{C}\simeq  \frac{3 \hbar}{8\pi\nu\tau^2}\left[-\frac{1}{D Q^2+\tau_\phi^{-1}}-\frac{1}{D Q^2+\tau_\phi^{-1}+\tau_v^{-1}}\right].
\end{equation}
In the derivation of Eqs.~(\ref{eq:cav1}) and ~(\ref{eq:cav2}) we have included the phase relaxation time $\tau_\phi$ and exploited $D Q^2\tau_0\ll 1$.

The first term in the square brackets of Eqs.~(\ref{eq:cav1}) and ~(\ref{eq:cav2}) is large at $Q\to 0$ irrespective of $\epsilon_F/M$, and originates from the spin-singlet Cooperon mode in Eq.~(\ref{eq:g1}).
Its negative sign means that it makes a contribution towards WAL.

Equation~(\ref{eq:cav1}) displays a competition between WL and WAL, which is no different from that found in an ordinary metal with spin-orbit interactions. 
WL terms originate from the three spin triplet modes of Eq.~(\ref{eq:trip}).
WL prevails if $\tau_\phi^{-1}\gg\tau_s^{-1}$, whereas WAL rules if $\tau_{\phi}^{-1}\ll\tau_s^{-1}$.

Equation~(\ref{eq:cav2}) unveils two different regimes of WAL.
On one hand, if $\tau_\phi^{-1}\gg\tau_v^{-1}$, the spin-singlet Cooperon mode of Eq.~(\ref{eq:g2}) makes a contribution to $\delta\sigma$ that equals that of Eq.~(\ref{eq:g1}).
In this limit, quantum interference can be interpreted as coming from two identical and nearly-decoupled Dirac valleys.
On the other hand, if $\tau_\phi^{-1}\ll\tau_v^{-1}$, the contribution from Eq.~(\ref{eq:g2}) becomes relatively unimportant and the magnitude of WAL is halved.    
In other words, when the intervalley transition rate induced by the mass term $M\tau^z$ is fast compared to the phase relaxation rate, the two valleys contribute as one.
This is quite different from graphene, where strong intervalley scattering changes WAL into WL.\cite{mccann2006}
The underlying reason for such a qualitative difference is that in graphene a gapless valley-singlet mode is responsible for producing WL, whereas in a Weyl semimetal the valley-singlet Cooperons are strongly gapped by the $k_z$ band dispersion.

Substituting Eqs.~(\ref{eq:cav1}) and ~(\ref{eq:cav2}) in Eq.~(\ref{eq:ds2}) and doing the $Q$-integral, we arrive at 
\begin{align}
\label{eq:res_bulk}
&\delta G\simeq\alpha\, G_q\ln(\tau_\phi/\tau)\nonumber\\
&\alpha=\left\{\begin{array}{ccc} -1 & {\rm if } & \tau_\phi\ll\tau_s\\
				 1/2 & {\rm if }& \tau_\phi\gg(\tau_v,\tau_s)\\   
                                 1 & {\rm if } & \tau_\phi\ll\tau_v,
\end{array}\right.
\end{align}
where $\delta G\equiv W\delta\sigma$ is the quantum interference correction to {\em conductance} and 
\begin{equation}
G_q\equiv e^2/(2\pi^2 \hbar)
\end{equation}
 is a universal conductance unit.
In the derivation of Eq.~(\ref{eq:res_bulk}) we have used Eq.~(\ref{eq:ds2}). 
The reason why $\alpha=1/2$ when $\tau_\phi\gg(\tau_v,\tau_s)$ is that in such regime there is only one gapless Cooperon mode (hence $|\alpha|=1/2$), which is a spin-singlet (hence $\alpha=|\alpha|$).

While Eq.~(\ref{eq:res_bulk}) is valid in absence of external magnetic fields, the magnetoconductance $\Delta G(H)\equiv G(H)-G(0)\simeq\delta G(H)-\delta G(0)$ can be easily obtained from Eq.~(\ref{eq:res_bulk}) for $H$ perpendicular to the TI thin film. 
The replacement of $\int d^2 Q$ by an appropriate sum over Landau levels\cite{hikami1980} results in
\begin{align}
\label{eq:magres_bulk}
&\Delta G\simeq\alpha\, G_q f(H_\phi/H)\nonumber\\
&\alpha=\left\{\begin{array}{ccc} -1 & {\rm if } & \tau_H\ll\tau_s\\
                                  1/2 & {\rm if } & \tau_H\gg(\tau_v,\tau_s)\\                                   
                                  1 & {\rm if } & \tau_H\ll\tau_v,
\end{array}\right.
\end{align}
where $f(z)\equiv\ln z-\psi(1/2+z)$ with asymptotes $f(z)\propto z^{-2}$ for $z\gg 1$ and $f(z)\propto\ln(1/z)$ for $z\ll 1$, 
$\psi$ is the digamma function, 
\begin{equation}
\tau_H^{-1}\equiv \tau_\phi^{-1}+2 e D H/\hbar\,\,\mbox{  and   }\,\, H_\phi\equiv\hbar/(4 e D \tau_\phi).
\end{equation}

Three conclusions of experimental relevance can be extracted from Eqs.~(\ref{eq:res_bulk}) and ~(\ref{eq:magres_bulk}), which apply when highest occupied electronic states are all located near the $\Gamma$ point.
First, bulk TI bands can display $\alpha=-1$ (WL) as long as the chemical potential is sufficiently close to the bottom of the bulk conduction band.
Second, bulk TI bands can produce $\alpha=1$ when $\epsilon_F/M$ is sufficiently large.
Third, when $(\epsilon_F-M)/M$ is neither large nor small, $\alpha=1/2$ ensues; this is the conventional result expected for ordinary conducting thin films with strong spin-orbit coupling, and is the one that has been often presumed in experiments on TI films.\cite{chen2010,checkelsky2011,wang2011,he2011,chen2011,steinberg2011}
At $\tau_H\simeq\tau_s$ there is a crossover between $\alpha=-1$ and $\alpha=1/2$; likewise, at $\tau_H\simeq\tau_v$ there is a crossover between $\alpha=1/2$ and $\alpha=1$.

The particular expressions for $\tau_s$ and $\tau_v$ in Eqs.~(\ref{eq:tau_s}) and (\ref{eq:tau_v}) rely on our assumption of $V_{\rm dis}\propto {\bf 1}_4$. 
Spin-orbit coupled impurities and/or atomically sharp disorder potentials would induce additional spin- and valley-flip processes, whose rates $\tau_{sf}^{-1}$ and $\tau_{vf}^{-1}$ would have to be incorporated via $\tau_s^{-1}\to\tau_s^{-1}+\tau_{sf}^{-1}$ and $\tau_v^{-1}\to\tau_v^{-1}+\tau_{vf}^{-1}$.
If $\tau_{vf}^{-1}$ and $\tau_{sf}^{-1}$ are strong enough and insensitive to the value of $\epsilon_F/M$, then the only surviving regime of interference corrections is the conventional $\alpha=1/2$.

The conventional $\alpha=1/2$ can be found in a wide range of parameter space at low temperatures, whereas the unconventional $\alpha=-1$ and $\alpha=1$ emerge in the relatively narrow regimes $\tau\ll\tau_H\ll\tau_s$ and $\tau\ll\tau_H\ll\tau_v$ (respectively).
How accessible are these unconventional regimes?
Suppose $M\simeq 150 {\rm meV}$, $v\simeq 5\times 10^5 {\rm m/s}$ and a bulk carrier density of $n\simeq 3\times 10^{18} {\rm cm}^{-3}$.
This situation corresponds to having a small bulk Fermi surface.
Then, it follows that $\alpha\simeq -1$ for a fairly wide range of magnetic fields ($l_H/(12 l)\ll 1$, where $l_H\equiv(D \tau_H)^{1/2}$). 
The limit $\alpha\to 1$ is not accesible in this regime.
Instead, $\alpha\simeq 1$ should be accessible in (i) Weyl semimetals or in TIs with very narrow bandgaps, (ii) in TIs with large bandgap but $M({\bf k}_F)\simeq 0$. 
For the latter case it must be kept in mind that in the absence of spherical symmetry $M({\bf k}_F)$ is not constant on the Fermi surface.
Suppose $M\simeq 5 {\rm meV}$ and a bulk carrier density of $\simeq 2\times 10^{18} {\rm cm}^{-3}$.
Then, $\alpha\simeq 1$ in the range of fields corresponding to $l_H/(10 l)\ll 1$.
For typical thin films, the requirements for $\alpha=\pm 1$ are compatible with $k_F l\gg 1$.

Materials like BiTl(S$_{1-\delta}$Se$_\delta$)$_2$, where controlled changes of $\delta$ can tune $M$ from 0 to large values,\cite{xu2011}  appear to be good candidates to observe crossovers between different regimes of magnetoresistance in Eq.~(\ref{eq:magres_bulk}).

Our analysis has thus far neglected surface states of the TI, which can also contribute to the measured magnetoresistance.
It can be argued that surface states are unimportant and Eq.~(\ref{eq:magres_bulk}) suffices in trivial insulators described by Eq.~(\ref{eq:model_b}), as well as in time-reversal-invariant Weyl semimetals and in TIs with very small bulk bandgaps ($\lesssim\hbar/\tau_0$).
In contrast, when the surface states of the TI are robust, Eq.~(\ref{eq:magres_bulk}) is incomplete and must be generalized. 
Such generalization is the subject for the rest of this paper. 

\section{Quantum Corrections to Conductivity from Coupled Bulk and Surface States}

In this section we consider the combined bulk-surface contribution to $\delta\sigma$ in 3D TIs with relatively large bandgaps.
We concentrate on a particular setup that consists of a TI thin film gated on one surface. 
The gate voltage can produce a depletion layer that spatially separates bulk and surface carriers (Fig.~\ref{fig:dep}), and carriers tunnel back and forth across the depletion layer. 
We assume the bulk-surface tunneling rate to be much smaller than the elastic scattering time on either side of the depletion layer, so that electrons scatter many times within the bulk (surface) before tunneling to the surface (bulk).
This assumption is experimentally realistic,  and it simplifies the microscopic theory of this section considerably.

\subsection{Single isolated TI surface}

As a preliminary step, we recall the expression for $\delta\sigma$ on a single TI surface that is decoupled from the bulk.
Taking $\epsilon_{Fs} \tau\gg 1$, where $\epsilon_{Fs}$ is the Fermi energy measured from the Dirac point of the surface states, one arrives\cite{tkachov2011,lu2011a} at
\begin{equation}
\label{eq:res_s}
\Delta G/G_q=(1/2) f(H_\phi/H)
\end{equation}
for any $\tau_H$. 
The prefactor $1/2$ is consistent with having a gapless spin-singlet Cooperon (the spin-triplet Cooperons have large gaps due to the strong spin-momentum coupling on the surface).

\subsection{Two coupled 2D layers without spin-orbit coupling}

As another preliminary step, here we compute $\delta\sigma$ for two ordinary metallic 2D layers separated by a tunnel barrier. 
In a double layer system, the current flowing in layer $i$ can be written as ${\bf j}_i=\sum_j \sigma_{i j} {\bf E}_j$, where ${\bf E}_j$ is the electric field in layer $j$.
For concreteness we take ${\bf E}_1={\bf E}_2\equiv {\bf E}$, so that the measured current is ${\bf j}={\bf j}_1+{\bf j}_2=\sigma{\bf E}$ with $\sigma=\sum_{i j}\sigma_{i j}$.
Consequently, the quantum corrections to conductivity are $\delta\sigma=\sum_{i j}\delta\sigma_{i j}$.
The goal of this section is to compute $\delta\sigma$ from microscopic theory.

\begin{figure}
\begin{center}
\includegraphics[scale=0.40]{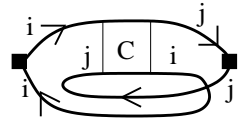}
\caption{Diagrammatic representation for $\delta\sigma_{i j}$, where $i$ and $j$ are layer indices. For 2D layers without spin-orbit coupling, the Cooperon matrix elements are fully characterized by layer indices. The velocity operator is diagonal in the layer index; therefore, the Cooperons $C^{1 1}_{2 2}$ and $C^{2 2}_{1 1}$ do not enter in the expression for $\delta\sigma_{i j}$.}
\label{fig:dsij}
\end{center}
\end{figure}

\begin{figure}
\begin{center}
\includegraphics[scale=0.35]{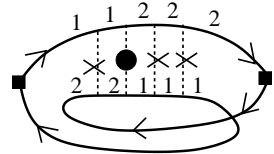}
\caption{Typical microscopic process that gives rise to $\delta\sigma_{1 2}$. It can be neglected when the intralayer disorder potentials in the two layers are uncorrelated.}
\label{fig:ds12}
\end{center}
\end{figure}

The interference correction $\delta\sigma_{i j}$ has the diagrammatic representation shown in Fig.~\ref{fig:dsij}.
Because the velocity operator is diagonal in the layer index, the only Cooperons that enter in the conductivity are $C^{i j}_{j i}$, with $i,j\in\{1,2\}$.
In particular $\delta\sigma_{i i}$ involves intralayer Cooperons $C^{i i}_{i i}$, whereas $\delta\sigma_{1 2}$ and $\delta\sigma_{2 1}$ involve interlayer Cooperons $C^{1 2}_{2 1}$ and $C^{2 1}_{1 2}$ (Fig.~\ref{fig:ds12}).
Assuming that disorder potentials in the two layers are uncorrelated, $C^{i j}_{j i}=0$ for $i\neq j$.
This is a reasonable assumption when electrons in the two layers scatter off different sets of impurities.
Hence, we are left with $\delta\sigma=\sum_i\delta\sigma_{i i}$.
From here on we simplify the notation via $C^{i i}_{i i}\equiv C_i$.

When evaluating $\delta\sigma_{i i}$ we will neglect spin-orbit interactions; however, the main lessons learned in this subsection will be transferrable to the spin-orbit coupled case studied in the next subsection.
In absence of interlayer coupling, a standard calculation yields
\begin{equation}
\label{eq:dnotu}
\delta\sigma_{i i}^{(0)}\simeq -4 \frac{e^2}{\hbar^2} \nu_i D_i  \tau_{d i}^2\int_{\bf Q} C_i^{(0)}({\bf Q}),
\end{equation}
where $\int_{\bf Q}\equiv \int d^2 Q/(2\pi)^2$, an extra factor of $2$ is due to spin degeneracy, $\tau_{d i}$ is the scattering time in layer $i$ due to elastic impurities (we assume purely s-wave scattering, so that there is no difference between the transport scattering time and the quantum lifetime), $\nu_i$ is the density of states per unit area in layer $i$ and 
\begin{equation}
C_i^{(0)} ({\bf Q})=\frac{\hbar}{2\pi \nu_i \tau_{d i}^2} \frac{1}{D_i Q^2+\tau_{\phi i}^{-1}}
\end{equation}
is the Cooperon for an isolated layer.
In presence of interlayer tunneling, $C_i^{(0)}$ in Eq.~(\ref{eq:dnotu}) is replaced by $C_i$:
\begin{equation}
\label{eq:dd}
\delta\sigma_{i i}\simeq -4 \frac{e^2}{\hbar^2} \nu_i D_i \tau_{d i}^2\int_{\bf Q}C_i({\bf Q}),
\end{equation}
in whose prefactor we have neglected terms containing the ratio between the tunneling rate and the elastic scattering rate.
\begin{figure}[t]
\begin{center}
\includegraphics[scale=0.35]{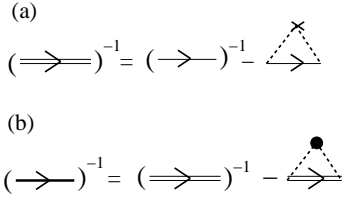}
\caption{Single-particle Green's functions. (a) Dressing of Bloch states due to intralayer impurity scattering. (b) Dressing of disorder-averaged Green's functions due to interlayer tunneling. The tunneling amplitude is regarded as a random variable.}
\label{fig:gfig}
\end{center}
\end{figure}

In order to compute $C_i$, we recognize that the influence of interlayer coupling occurs at two different levels.
On one hand, it modifies the single-particle Green's function for each layer (Fig.~\ref{fig:gfig}).
Because the thickness of the depletion layer typically shows microscopic variations within the same film as well as from sample to sample,
the interlayer tunneling amplitude can be regarded as a random variable.
Consequently, the change in the ensemble-averaged Green's function due to tunneling can be captured via $\tau_{d i}^{-1}\to \tau_{d i}^{-1}+\tau_{t i}^{-1}$, 
where 
\begin{equation}
\tau_{t i}^{-1}=(2\pi/\hbar) \langle|t|^2\rangle S\, \nu_j
\end{equation} 
is the tunneling rate from layer $i$ onto layer $j\neq i$, $\langle|t|^2\rangle$ is the averaged square of the tunneling matrix element and $S$ is the layer area.
Note that $\langle|t|^2\rangle$ scales like $S^{-1}$, so that $\tau_{t i}^{-1}$ is independent of the layer area.

\begin{figure}
\begin{center}
\includegraphics[scale=0.4]{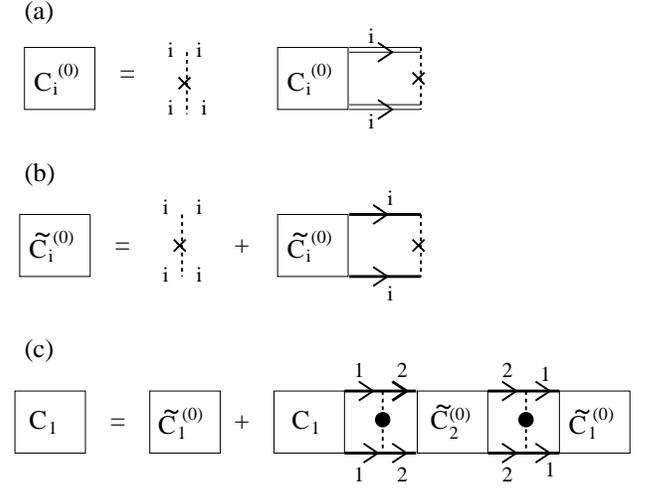}
\caption{(a) Cooperon $C_i^{(0)}$ without interlayer tunneling. (b) Partially dressed Cooperon $\tilde{C}_i^{(0)}$, where tunneling is included solely in the single-particle Green's functions. $\tilde{C}_i^{(0)}$ can be directly obtained from $C_i^{(0)}$ via $\tau_{\phi i}\to\tilde{\tau}_{\phi i}$. (c) Fully dressed Cooperon $C_i$, where tunneling is incorporated both in the single-particle Green's function and in the particle-particle correlations.}
\label{fig:cfig}
\end{center}
\end{figure}

\begin{figure}
\begin{center}
\includegraphics[scale=0.44]{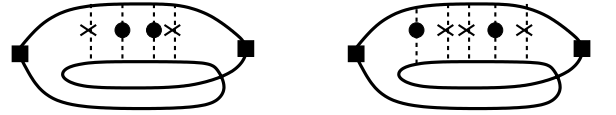}
\caption{Typical processes not included in Fig.~\ref{fig:cfig}, as they are subdominant for $\tau_{t i}\gg\tau_{d i}$.}
\label{fig:nfig}
\end{center}
\end{figure}

On the other hand, interlayer tunneling modifies particle-particle correlations that build up Cooperons.
An approximate diagrammatic expression for these correlations is shown in Fig.~\ref{fig:cfig}.
The equation of Fig.~\ref{fig:cfig}c can be solved in momentum space and it yields
\begin{equation}
\label{eq:cii}
C_i  =\frac{\hbar}{2\pi\nu_i\tau_{d i}^2}\frac{D_j Q^2+\tilde{\tau}_{\phi j}^{-1}}{(D_1 Q^2+\tilde{\tau}_{\phi 1}^{-1})(D_2 Q^2+\tilde{\tau}_{\phi 2}^{-1})-\tau_{t 1}^{-1}\tau_{t 2}^{-1}}
\end{equation}
for $j\neq i$.
In the derivation of Eq.~(\ref{eq:cii}) we have introduced
\begin{equation}
\label{eq:tau_tilde}
\tilde{\tau}_{\phi i}^{-1}\equiv\tau_{\phi i}^{-1}+\tau_{t i}^{-1} 
\end{equation}
as an effective phase relaxation rate that incorporates tunneling, and have used 
\begin{equation}
\int_{\bf k} G_i^R ({\bf k}) G^A_i(-{\bf k}+{\bf Q})\simeq \frac{2\pi \nu_i\tau_{d i} }{\hbar}\left(1-\frac{\tau_{d i}}{\tilde{\tau}_{\phi i}}-D_i Q^2 \tau_{d i}\right).\nonumber
\end{equation}
Microscopic processes depicted in Fig.~\ref{fig:cfig} leave out those in which two consecutive tunneling events occur without any intralayer scattering in between.
Likewise, they ignore electron trajectories in which a tunneling event precedes any intralayer scattering (Fig.~\ref{fig:nfig}).
These processes are relatively unimportant if $\tau_{t i}\gg \tau_{d i}$.
Not surprisingly, Eq.~(\ref{eq:cii}) arises in the coupled equations for the classical diffusive conductivity as well (see Appendix~\ref{sec:cond}).

It is convenient to rewrite $C_i$ in Eq.~(\ref{eq:cii}) as
\begin{equation}
\label{eq:q1}
C_i=\frac{\hbar}{2\pi\nu_i D_i \tau_{d i}^2}\left[\frac{A_i}{Q^2+q_a^2}+\frac{B_i}{Q^2+q_b^2}\right], 
\end{equation}
where
\begin{equation}
\label{eq:ab1}
2 q_{a (b)}^2=\frac{1}{\tilde{l}_{\phi 1}^{2}}+\frac{1}{\tilde{l}_{\phi 2}^{2}}\pm \sqrt{\left(\frac{1}{\tilde{l}_{\phi 1}^{2}}-\frac{1}{\tilde{l}_{\phi 2}^{2}}\right)^2+ \frac{4}{l_{t 1}^2 l_{t 2}^2}}
\end{equation}
and
\begin{equation}
\label{eq:ab2}
A_i = 1-B_i= (\tilde{l}_{\phi j}^{-2}-q_a^2)/(q_b^2-q_a^2)\,\,\mbox{ for $j\neq i$.}
\end{equation}
In Eq.~(\ref{eq:ab1}) we have defined $\tilde{l}_{\phi i}\equiv(D_i\tilde{\tau}_{\phi i})^{1/2}$ as an effective coherence length and $l_{t i}\equiv(D_i\tau_{t i})^{1/2}$ as the interlayer leakage length. 
Besides, $q_a^2 (q_b^2)$ gets the positive (negative) sign in front of the square root.
Combining Eq.~(\ref{eq:dd}) with Eq.~(\ref{eq:q1}) and using $A_1+A_2=B_1+B_2=1$, we get
\begin{equation}
\delta\sigma=\sum_i\delta\sigma_{i i}=-2 \frac{e^2}{\pi\hbar} \int_{\bf Q} \left[\frac{1}{Q^2+q_a^2}+\frac{1}{Q^2+q_b^2}\right].
\end{equation}
Therefore, the low-field magnetoconductance reads
\begin{equation}
\label{eq:Dii}
\Delta\sigma=\sum_i\Delta\sigma_{i i}=-G_q\left[ f\left(\frac{H_a}{H}\right)+ f\left(\frac{H_b}{H}\right) \right],
\end{equation}
where 
\begin{equation}
H_{a (b)}\equiv \hbar\, q_{a (b)}^2/(4 e).
\end{equation}
In the limit of very strong tunneling ($\tau_{t i}/\tau_{\phi i}\to 0$), Eq.~(\ref{eq:Dii}) becomes $\Delta\sigma\simeq -G_q f(H_b/H)$, as though there was a single layer.
In the limit of very weak tunneling ($\tau_{t i}/\tau_{\phi i}\to\infty$), $\Delta\sigma$ is the sum of contributions from two independent films. 

It is helpful to understand the weak and strong coupling regimes in terms of measurable quantities like the interlayer conductance per square, 
\begin{equation}
g_t=(2\pi e^2/\hbar)\langle|t|^2\rangle S \nu_1\nu_2=\sigma_{D i}/l_{t i}^2,
\end{equation}
where $\sigma_{D i}$ is the Drude conductivity in layer $i$.
For simplicity suppose that $\tau_{\phi 1}\simeq\tau_{\phi 2}\equiv\tau_\phi$.
In this case the crossover from weak to strong tunneling occurs when 
\begin{equation}
\label{eq:cross}
\frac{1}{g_t l_\phi^2}\lesssim \frac{1}{\sigma_{D 1}}+\frac{1}{\sigma_{D 2}}\,\,\,\mbox{(crossover condition)},
\end{equation}
namely when the tunneling resistance for a square of area $l_\phi^2$ becomes smaller than the sum of the classical intralayer resistivities.
Let us define 
\begin{equation}
g_c^{-1}\equiv (\sigma_{D 1}^{-1}+\sigma_{D 2}^{-1}) l_\phi^2.
\end{equation}
If $g_t\ll g_c$, then $\Delta\sigma/G_q\simeq -2\ln(H/H_\phi)$ for $H\gg H_\phi$. 
If $g_t\gg g_c$, then $\Delta\sigma/G_q\simeq -\ln(H/H_\phi)$ for $H_\phi\ll H\ll H_\phi (g_t/g_c)$.
Thus changing the interlayer conductance results in a factor-of-two change for the magnitude of the WL correction.

\begin{figure}
\begin{center}
\includegraphics[scale=0.35]{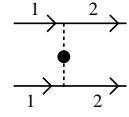}
\caption{Example of an interlayer scattering process that is allowed in multilayer systems. Its analog in multivalley semiconductors of Ref.~[\onlinecite{fukuyama1980}] is forbidden.}
\label{fig:fuku}
\end{center}
\end{figure}

Limits reminiscent of the above were first discussed in inversion layers of multivalley semiconductors like Si,\cite{fukuyama1980} where the role of the layers is played by different electron pockets in the Brillouin zone.
Similarities notwithstanding, there are clear differences between our microscopic theory and that of multivalley semiconductors. 
On one hand, the separation in momentum between valleys of Si prevents scattering processes such as the one in Fig.~\ref{fig:fuku}. 
These processes are not only allowed in our case, but also lead to the Cooperon dressing shown in Fig~\ref{fig:cfig}c. 
On the other hand, in our case the interlayer Cooperon vanishes due to uncorrelated disorder potentials in the two spatially separated layers. 
That is not the case in multivalley semiconductors, where both valleys scatter off the same set of real-space impurities and intervalley Cooperons contribute crucially to $\delta\sigma$. 

Finally, it should be mentioned that Eqs.~(\ref{eq:ab1}), (\ref{eq:ab2}) and (\ref{eq:Dii}) coincide with those derived by G. Bergmann,\cite{bergmann1989} who invoked macroscopic arguments based on the diffusion equation.
The microscopic theory of this subsection supports Bergmann's results, insofar as $\tau_{t i}\gg \tau_{d i}$ and the disorder potentials in the two layers are uncorrelated.
Incidentally, yet another way to arrive at the same results is unveiled in Appendix~\ref{sec:coupled_cooper}; this later method will prove convenient in the upcoming subsection.

\subsection{3D TI film with bulk-surface coupling}

We now consider a 3D TI film (Fig.~\ref{fig:dep}) with a gate electrode placed near its top surface. 

At the moment we neglect the bottom surface of the TI, which will be incorporated below.
For ease of notation we use subscript ``1'' to refer to ``bulk'', and subscript ``2'' to refer to ``top surface''. 
Like in the preceding subsection we assume bulk-surface disorder correlations to be negligible, so that the quantum corrections to conductance can be written as $\delta G=\delta G_{1 1}+\delta G_{2 2}=W \delta\sigma_{1 1}+\delta\sigma_{2 2}$.
$\delta G$ is approximately independent of the film thickness $W$ as long as $W\ll\tilde{l}_{\phi 1}$ , where $\tilde{l}_{\phi 1}$ was defined below Eq.~(\ref{eq:ab2}).

The goal of this subsection is to calculate $\delta G$ from microscopic theory.
Unlike in the previous subsection, here both ``layers'' are spin-orbit coupled.
We assume that tunneling events, albeit being time-reversal invariant, conserve neither spin nor orbital degrees of freedom.
Indeed, in a TI spin is not conserved for non-momentum-conserving tunneling.
Similarly, the orbital degree of freedom is not conserved due to broken inversion symmetry near the surface.

Let us begin with no tunneling.
On one hand, there are four surface Cooperon modes: one gapless spin-singlet mode and three spin-triplet modes with large ($\sim\tau_{d 2}^{-1}$) gaps.
On the other hand, there are sixteen bulk Cooperons, of which a spin-singlet mode (Eq.~(\ref{eq:g1})) is always gapless.
In addition, four of the bulk modes (the spin-singlet of Eq.~(\ref{eq:g2}) and three spin-triplets of Eq.~(\ref{eq:trip})) can be ``soft'' depending on $\epsilon_F/M$.
The rest of the bulk Cooperon modes have large gaps of order $\tau_{d 1}^{-1}$.

Let us now turn on tunneling.
Since $\tau_{t i}\gg \tau_{d i}$, we can limit ourselves to analyzing the effects of tunneling within the low-energy subspace formed by the soft Cooperons. 
If there are no magnetic impurities in the depletion layer, the total spin of the Cooperon is a good quantum number even in presence of tunneling. 
Accordingly tunneling does not mix spin-singlet modes with spin-triplet modes, and the full (dressed) Cooperons can also be classified into spin-singlets and a spin-triplet. 

In the regimes $\tau_{\phi 1}\ll\tau_s$ and $\tau_{\phi 1}\gg (\tau_s,\tau_v)$, tunneling dresses one soft spin-singlet Cooperon in the bulk with another soft spin-singlet Cooperon on the surface.
This dressing is completed as explained in Section IIIB: first by renormalizing the phase relaxation time  $\tau_{\phi i}\to \tilde{\tau}_{\phi i}$, 
and afterwards proceeding with the series expansion of Fig.~\ref{fig:cfig}c. 
All ``blocks'' appearing in this series expansion are spin-singlets.
When $\tau_{\phi 1}\ll\tau_s$, the soft spin-triplet Cooperons from the bulk are dressed simply through $\tau_{\phi 1}\to\tilde{\tau}_{\phi 1}$: they do not get appreciably admixed with the spin-triplet Cooperon on the surface because the latter has a large gap.

In the regime $\tau_{\phi 1}\ll\tau_v$, there are two gapless singlet Cooperons in the bulk, each of which can hybridize with the singlet gapless Cooperon on the surface.
For this situation, Fig.~\ref{fig:cfig}c does not capture all possible processes and the calculation from the previous subsection must be generalized; this generalization is carried out in Appendix~\ref{sec:coupled_cooper}.

With the above considerations in mind, we combine Eqs.~(\ref{eq:magres_bulk}) and ~(\ref{eq:res_s}) in order to obtain the total contribution to low-field magnetoconductance:
\begin{align}
\label{eq:res_tot}
& \frac{\Delta G}{G_q}\simeq\frac{1}{2}\left\{\begin{array}{ccc} 
f\left(\frac{H_a}{H}\right)+f\left(\frac{H_b}{H}\right)-3 f\left(\frac{\tilde{H}_{\phi 1}}{H}\right) &{\rm if } & \tilde{\tau}_H\ll\tau_s\\
f\left(\frac{H_a}{H}\right)+f\left(\frac{H_b}{H}\right) &{\rm if } & \tilde{\tau}_H\gg(\tau_s,\tau_v)\\
f\left(\frac{H_c}{H}\right)+f\left(\frac{H_d}{H}\right)+f\left(\frac{\tilde{H}_{\phi 1}}{H}\right) &{\rm if } &\tilde{\tau}_H\ll\tau_v,
\end{array}\right.
\end{align}
where $H_l=\hbar\, q_l^2/(4 e)$ for $l\in\{a,b,c,d\}$,
\begin{equation}
\tilde{H}_{\phi 1}\equiv \hbar/(4 e D_1 \tilde{\tau}_{\phi 1}),\,\,\,{\rm and}\,\,\, \tilde{\tau}_H^{-1}\equiv \tilde{\tau}_{\phi 1}^{-1}+ 2 e D_1 H/\hbar. 
\end{equation}
Note that the effective phase relaxation rate increases linearly with the bulk-to-surface tunneling rate (c.f. Eq.~(\ref{eq:tau_tilde})).
The characteristic momenta $q_{a (b)}$ have been introduced earlier in Eq.~(\ref{eq:ab1}). 
The additional momenta $q_{c(d)}$ are identical to $q_{a (b)}$, except for $\tau_{t 2}^{-1}\to 2\,\tau_{t 2}^{-1}$. 
The reason for this difference is that the surface Cooperon can decay into two gapless bulk Cooperons when $\tau_{\phi 1}\ll\tau_v$.

The first line of Eq.~(\ref{eq:res_tot}) displays a competition between WL and WAL, and suggests that it is possible to induce a WAL-to-WL transition with a varying gate voltage.
In the weak tunneling regime WL prevails, whereas in the strong tunneling regime WAL takes over.
Similarly, a gate voltage can induce transitions between three different WAL coefficients: $\alpha\in(1/2,1)$ in the second line, and $\alpha\in(1/2,3/2)$ in the third line. 
The second line of Eq.~(\ref{eq:res_tot}) describes quantum corrections as if they originated from two independent thin films with mixed bulk-surface character;
indeed, universal results expected for the simplectic symmetry class are recovered when the effective phase relaxation times become the longest timescales of the problem.
Some simple limiting cases of Eq.~(\ref{eq:res_tot}) are discussed in Appendix~\ref{sec:special}. 

Thus far we have considered the coupling between the bulk and {\em one} (the top) surface of the TI film. 
As a consequence, Eq.~(\ref{eq:res_tot}) applies to a TI film only if the phase relaxation time of the bottom surface (adjacent to the substrate) is short compared to other phase relaxation and tunneling times in the problem.
This condition is likely not met in some recent experiments,\cite{checkelsky2011,chen2011} which report on independent contributions from both surfaces.
Partly motivated by these experiments, we now generalize Eq.~(\ref{eq:res_tot}) so as to capture two surfaces, each coupled to bulk states.

We consider the scenario depicted in Fig.~\ref{fig:dep}, where the bottom surface contains bulk carriers.
Since there is no depletion layer near $z=W$, we assume that the bulk-surface tunneling rate therein is strong compared to the phase relaxation rate, yet weak compared to disorder scattering rate.
Hence we describe the hybrid of bottom surface and bulk states via Eq.~(\ref{eq:res_tot2}), and thereafter couple this hybrid to the top surface along the lines of Eq.~(\ref{eq:res_tot}).
The resulting expression for $\Delta G$ can be approximated as 
\begin{equation}
\label{eq:res_tot5}
\frac{\Delta G}{G_q}\simeq\frac{1}{2} f\left(\frac{H'_a}{H}\right)+\frac{1}{2}f\left(\frac{H'_b}{H}\right),
\end{equation}
where $H'_{a(b)}\equiv\hbar (q'_{a(b)})^2/(4 e)$.
The characteristic momenta $q'_a$ and $q'_b$ obey Eq.~(\ref{eq:ab1}), where  ``1'' labels the top surface and ``2'' labels a hybrid between the bottom surface and the bulk.

Notably, Eq.~(\ref{eq:res_tot5}) implies that WL is no longer possible once the bottom surface is strongly coupled to bulk states. 
Instead, conventional WAL ensues with $\alpha\in(1/2,1)$.
This observation not only sheds light on why current experiments see no indication for WL, but it also gives insight as to how WL could be observed in TI films.

A possible strategy is to degrade the surfaces, e.g. by depositing magnetic impurities on them, and decoupling them from the bulk by double-sided gating. 
One may expect WL even if only the top surface is decoupled, while the (degraded) bottom surface is in contact with the bulk.
In this case, Eq.~(\ref{eq:res_tot}) reduces to Eq.~(\ref{eq:magres_bulk}) derived for the sole bulk conduction, with the replacement $\tau_{\phi 1}^{-1}\to\tau_{\phi 1}^{-1}+\tau_{t 3}^{-1}$, where $\tau_{t 3}$ is the tunneling rate of electrons from bulk to the bottom surface.
If the film is thick enough, then $\tau_{t 3}^{-1}$ may become sufficiently small to provide some dynamic range for observing WL behavior.
This same strategy can also facilitate the observation of WAL with $\alpha>1$.

\subsection{Estimates for the bulk-surface coupling} 

\begin{figure}
\begin{center}
\includegraphics[scale=0.35]{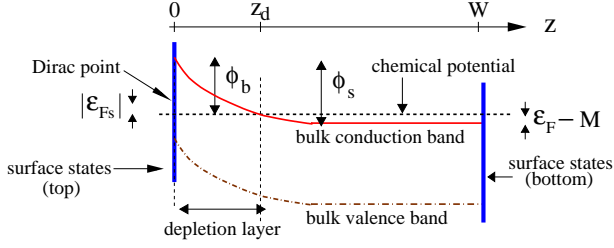}
\caption{Schematic energy band profile for a gated 3D TI thin film. $z=0$ corresponds to the top surface of the device, immediately under the gate. 
$z=W$ corresponds to the bottom (ungated) surface.
The vertical (blue) solid lines at $z=0, W$ are surface states. 
The curved solid (red) line is the bulk conduction band, and the dot-dashed (brown) curve is the bulk valence band. 
The chemical potential is depicted by a horizontal dashed line. 
$z_d$ is the thickness of the depletion layer, where neither bulk nor surface carriers are present. 
$\epsilon_{F s}$ is the Fermi energy of the surface states measured from the Dirac point ($\epsilon_{Fs}<0$ in this figure). 
$\epsilon_F$ is the Fermi energy of the bulk states, measured with respect to the midgap point.  
}
\label{fig:dep}
\end{center}
\end{figure}

This subsection is devoted to an approximate electrostatic and quantum mechanical analysis of the depletion layer in a TI film, which will result in quantitative estimates for the bulk-surface coupling.

For a TI with an $n$-doped bulk, a negative charge per unit area $(-Q_g)$ placed at the gate repels electrons from bulk bands at $z=0$ as well as from the surface states at $z=0$.
This leaves a positive net charge on the top surface, which is equivalent to a downward shift in the local chemical potential at $z=0$:  $\Delta\mu_s=\epsilon_F-\epsilon_{F s}$.
Recall that $\epsilon_{Fs}$ is the Fermi energy of the surface states measured from the Dirac point (for simplicity the Dirac point is assumed to be in the middle of the bandgap at $z=0$) and $\epsilon_F$ is the Fermi energy of the bulk states measured from the middle of the bandgap. 
Since the chemical potential deep inside the bulk must be unaffected by the gate, $\Delta\mu_s\neq 0$ implies a band bending of magnitude $\phi_s=\Delta\mu_s$ near the gated surface (Fig.~\ref{fig:dep}).

When $\Delta\mu_s>(\epsilon_F-M)$ there are no bulk carriers left at $z=0$ and a depletion layer appears at $z\in(0,z_d)$, where $z_d$ is determined below.
For each value of $Q_g$, $\Delta\mu_s$ (or equivalently $\epsilon_{F s})$ can be uniquely determined from the overall neutrality condition $Q_s+Q_d=Q_g$, where $Q_s$ is the positive net charge induced on the surface, and $Q_d$ is the positive net charge in the depletion layer. 

In the depletion approximation\cite{sze2002} one has $Q_d\simeq n z_d$, where $n$ (c.f. Eq.~(\ref{eq:n})) is equal to the density of charged donors in the depleted region.
The electrostatic energy profile in the depleted region then obeys
\begin{equation}
\label{eq:fi}
\phi(z)=\phi_b-\frac{1}{2}\frac{e^2 n}{\kappa}(z-z_d)^2,
\end{equation}
where $\phi_b\equiv\phi_s-(\epsilon_F-M)=M-\epsilon_{F s}$, $\kappa$ is the static dielectric constant and 
\begin{equation}
\label{eq:depl}
z_d=\sqrt{\frac{2\kappa \phi_b}{e^2 n}}.
\end{equation} 
In the derivation of Eq.~(\ref{eq:fi}) we have assumed that the electric field vanishes at $z=z_d$, which is accurate within a screening radius. 
As the gate voltage is made more negative, the maximum width of the depletion layer ($z_d^{\rm max}$) is achieved when $\phi_b\simeq 2 M$.
For $\phi_b>2 M$, the bulk bands get inverted at $z=0$ and $z_d$ saturates.
We estimate $z_d^{\rm max}\simeq 20 {\rm nm}$ for some typical parameter values ($M=150 {\rm meV}$, $n\simeq 4\times 10^{18} {\rm cm}^{-3}$, $\kappa=50$).

Once the electrostatic profile of the TI film is characterized, we can analyze the quantum mechanical tunneling of electrons across the depletion layer. 
The tunneling conductance per unit area is roughly
\begin{equation}
\label{eq:kappa}
g_t\sim (e^2/h) \lambda_F^{-2} \exp(-2\chi),
\end{equation}
where $\lambda_F$ is the smallest between bulk and surface Fermi wavelengths, and 
\begin{equation}
\label{eq:kappa2}
\chi\simeq \int_0^{z_d} dz\frac{\phi_b-\phi(z)}{\hbar v}\simeq\frac{1}{6}\frac{e^2 n z_d^3}{\kappa\,\hbar v}.
\end{equation} 
In Eq.~(\ref{eq:kappa2}) we have ignored effective mass and Fermi velocity mismatches across the depletion layer.
The WKB exponent $\chi$ can be tuned by a gate voltage: as $z_d$ varies from $0$ to $z_d^{\rm max}$, $\chi$ goes from $0$ to $\simeq 6$. 

Drawing from the previous subsection (c.f. Eq.~(\ref{eq:cross})), the crossover from weak to strong bulk-surface coupling occurs when
\begin{equation}
\label{eq:cross1}
\frac{1}{g_t l_\phi^2}\lesssim\frac{1}{\sigma_{D 1} W}+\frac{1}{\sigma_{D 2}}\simeq \frac{1}{\sigma_{D 2}},
\end{equation}
where in the second equality we have assumed that $\sigma_{D 1} W\gg\sigma_{D 2}$.
This is a good assumption provided that (i) the bulk mean free path is of the same order as the surface mean free path, and (ii) $k_F W\gg 1$.
Plugging Eq.~(\ref{eq:kappa}) in Eq.~(\ref{eq:cross1}), the latter becomes
\begin{equation}
\label{eq:cross2}
\frac{l_\phi}{\lambda_F}\gtrsim (k_{Fs} l_2)^{1/2} \exp(\chi),
\end{equation} 
where $k_{Fs}=|\epsilon_{F s}|/{\hbar v}$ is the Fermi wave vector for the surface states and we have used $\sigma_{D 2}\sim (e^2/h) k_{F s} l_2$.
 
When $z_d=z_d^{\rm max}$, the right hand side of Eq.~(\ref{eq:cross2}) reaches $\simeq 1000$, which exceeds the typical $l_\phi/\lambda_F$ in TI thin films by at least an order of magnitude.
Therefore, when the depletion layer has its maximum width, the top surface and the bulk of the TI film can be regarded as weakly coupled.
This state of affairs changes rapidly when the depletion layer is made thinner by a gate voltage.
For instance, when $z_d=z_d^{\rm max}/\sqrt2$, the right hand side of Eq.~(\ref{eq:cross2}) equals $\simeq 30$, which is comparable to the typical $l_\phi/\lambda_F$.
Further slight reductions in $z_d$ can subsequently drive the film into a regime of strong bulk-surface coupling.
These estimates justify the interpretation of experimental data given in e.g. Ref.~[\onlinecite{steinberg2011}].

\section{Summary and conclusions}

We have completed a theoretical study of low-field magnetoresistance in electrostatically gated 3D TI films.
The concise analytical expressions presented here [Eqs.~(\ref{eq:magres_bulk}),~(\ref{eq:res_tot}) and (\ref{eq:res_tot5})] may shed light on the quantum magnetoresistance of TIs, Weyl semimetals, as well as some topologically trivial materials.
Only magnetic fields that are perpendicular to the TI thin film have been considered in this work; for in-plane fields and small bulk bandgaps, quantum interference contributions might be masked by classical magnetoresistance anomalies.\cite{son2012}
 
A number of predictions from this work have not been articulated in previous studies and await experimental confirmation.  
For instance, we find that TI thin films with low bulk doping may exhibit weak localization (WL) or negative magnetoresistance, instead of the often presumed weak antilocalization (WAL) or positive magnetoresistance.
Admittedly, the parameter space for WL is relatively narrow, and vanishes when either surface of the TI film is strongly coupled to bulk states.
However, WL may be experimentally accessible in thicker films, or in thin films where the surfaces have short phase relaxation times. 
Under these conditions, a gate can induce a crossover between WL and WAL.
On a separate note, we find that the ``universal'' prefactor for WAL varies depending on the bandgap of the TI, on the bulk doping concentration, on the phase relaxation times, and on the applied gate voltage.
 
The results from this work are applicable to conducting yet lighly doped TIs, with thicknesses ranging between the bulk transport mean free path and the bulk phase relaxation length.
It may be useful to find out how the results derived here change in highly doped TIs containing additional electrons pockets away from the $\Gamma$ point.
Likewise, it may be helpful to extend our results to thinner films.
Other potentially interesting tasks involve investigating universal conductance fluctuations and determining the influence of electron-electron interactions in the magnetoresistance of doped TI films.

\acknowledgments
This research has been financially supported by a fellowship from Yale University (I.G.), and by NSF DMR Grant No. 0906498 (L.G.).
L.G. thanks Pablo Jarillo-Herrero for a discussion that initiated the present work, I.G. thanks Ewelina Hankiewicz for an informative conversation, and both authors thank Aharon Kapitulnik for bringing Ref.~[\onlinecite{kapitulnik2012}] to their attention.

\appendix
\begin{widetext}
\section{Renormalized velocity operator}
\label{sec:ren}

The velocity operators appearing in the expressions for $\sigma_D$ and $\delta\sigma$ (c.f. Sec. IIB) must be renormalized with ladder diagrams containing impurity scattering. 
The Dyson equation for the renormalized velocity operator is (Fig.~\ref{fig:vertex})
\begin{equation}
\label{eq:vd}
\tilde{{\bf v}}_{\alpha\beta}({\bf k})= {\bf v}_{\alpha\beta}({\bf k})+u_0\sum_{\alpha,\beta\in\{1,2\}}\int_{{\bf k}'}\langle\alpha {\bf k}|\alpha' {\bf k}'\rangle\langle\beta'{\bf k}'|\beta{\bf k}\rangle G^A({\bf k}') G^R({\bf k}') \tilde{{\bf v}}_{\alpha'\beta'}({\bf k}'),
\end{equation}
where ${\bf v}_{\alpha\beta}({\bf k})=\delta_{\alpha\beta} \hbar v^2 {\bf k}/E_k$ is a matrix element for the bare velocity operator.
We solve Eq.~(\ref{eq:vd}) by guessing a solution of the form
\begin{equation}
\label{eq:guess}
\tilde{{\bf v}}_{\alpha\beta}({\bf k})=\gamma_k {\bf k}\delta_{\alpha\beta},
\end{equation}
where $\gamma_k$ is a scalar that depends on $|{\bf k}|$ but not $\hat{\bf k}$.
Although it is {\em a priori} not obvious that the renormalized velocity operator should be diagonal in the band indices, substituting Eq.~(\ref{eq:guess}) in Eq.~(\ref{eq:vd}) and using Eq.~(\ref{eq:eigenstates}) we find that $\tilde{{\bf v}}_{\alpha\beta}({\bf k})\propto\delta_{\alpha\beta}$ is indeed appropriate provided that
\begin{equation}
\gamma_k=\frac{\hbar v^2}{E_k}\frac{\tau}{\tau_0}.
\end{equation}
Here 
\begin{equation}
\frac{\hbar}{\tau}=2\pi\nu u_0\int\frac{d\Omega_{{\bf k}'}}{4\pi}\sum_{\alpha'}|\langle\alpha {\bf k}_F|\alpha' {\bf k}'_F\rangle|^2 (1-\hat{\bf k}_F\cdot\hat{\bf k}'_F)
\end{equation}
is the transport scattering time. 
Therefore, the final result for the renormalized velocity is $\tilde{{\bf v}}_{\alpha\beta}({\bf k})={\bf v}_{\alpha\beta}({\bf k}) (\tau/\tau_0)$.

\begin{figure}[h]
\begin{center}
\includegraphics[scale=0.35]{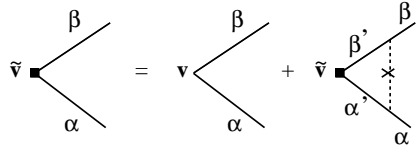}
\caption{Impurity vertex corrections for the velocity operator
}
\label{fig:vertex}
\end{center}
\end{figure}

\section{Evaluation of $\delta\sigma_2$ in some simple cases}
\label{sec:ds2}

The expression for $\delta\sigma_2$ (depicted in Fig.~\ref{fig:cofig}c) reads
\begin{equation}
\label{eq:nasty}
\delta\sigma_2\simeq -2\frac{e^2\hbar}{2\pi}\int_{{\bf k}, {\bf k}'} \tilde{v}^x({\bf k}) \tilde{v}^x({\bf k}') G^A({\bf k}) G^A({\bf k}') G^A(-{\bf k})
G^A(-{\bf k}') G^R(-{\bf k}') G^R({\bf k})\sum_{\alpha\beta\alpha'\beta'}\Gamma^{\alpha\alpha'}_{\beta'\beta}({\bf k},-{\bf k}',0)
\frac{1}{W}\int\frac{d^2 Q}{(2\pi)^2} C^{\beta\alpha}_{\alpha'\beta'}({\bf k},{\bf k}',{\bf Q}),
\end{equation}
 where the overall factor of two stems from the fact that the two diagrams in Fig.~\ref{fig:cofig}c give identical contribution, and the band indices $\alpha,\beta$ etc. are summed over $1,2$.
For generic $(\epsilon_F-M)/M$, the calculation of $\delta\sigma_2$ is cumbersome.
Here we focus on two simple limits that are of interest: $(\epsilon_F-M)/M\ll 1$ and $(\epsilon_F-M)/M\gg 1$.

When $(\epsilon_F-M)/M\ll 1$, the momentum dependence of $|\alpha {\bf k}_F\rangle$ is negligible. 
Consequently, $C^{\beta\alpha}_{\alpha'\beta'}({\bf k}_F,{\bf k}_F',{\bf Q})$ and $\Gamma^{\alpha\alpha'}_{\beta'\beta}({\bf k}_F,-{\bf k}_F',0)$ become independent of ${\bf k}_F$ and ${\bf k}_F'$.
Since the matrix elements of the velocity operator are odd under ${\bf k}\to -{\bf k}$ and ${\bf k}'\to -{\bf k}'$, it is clear that 
\begin{equation}
\delta\sigma_2\simeq 0. 
\end{equation}
  
The limit of $(\epsilon_F-M)/M\gg 1$ is less trivial.
In this regime the Hamiltonian is approximately block diagonal both in absence and in presence of disorder,  because the disorder potential we take is spin- and orbital-indpendent.
Therefore we may focus on a $2\times 2$ Hamiltonian (describing a Weyl node of positive chirality),
\begin{equation}
\label{eq:h_simple}
h'({\bf k})=\hbar v {\bf k}\cdot{\boldsymbol \sigma}+V_0({\bf r}){\bf 1}_{2\times 2},
\end{equation}
where ${\bf k}=k(\sin\theta\cos\phi,\sin\theta\sin\phi,\cos\theta)$.
The result for $\delta\sigma$ obtained from such Hamiltonian needs to be multiplied by two at the end, as each block makes an equal contribution.
The eigenstates for $h'({\bf k})$ are $|+,{\bf k}\rangle=(\cos(\theta/2),\exp(i\phi)\sin(\theta/2))^T$ and $|-,{\bf k}\rangle=(\sin(\theta/2),-\exp(i\phi)\cos(\theta/2))^T$. 

One significant simplification from Eq.~(\ref{eq:h_simple}) is that there is only one band at the Fermi energy. 
This allows us to rewrite Eq.~(\ref{eq:nasty}) as
\begin{align} 
\label{eq:less_ugly}
\delta\sigma_2=& -4\frac{e^2\hbar^3}{2\pi} u_0\frac{\tau^2}{\tau_0^2}\left[\int\frac{dk k^2}{(2\pi)^2}\frac{k v^2}{E_k} (G^A)^2 G^R\right]^2\frac{1}{W}\int\frac{d^2 Q}{(2\pi)^2}\nonumber\\
&\times\int\frac{d\Omega_{{\bf k}}}{4\pi}\int\frac{d\Omega_{{\bf k}'}}{4\pi} \sin\theta\cos\phi\sin\theta'\cos\phi'\langle +,{\bf k}_F|+,{\bf k}_F'\rangle\langle +,-{\bf k}_F|+,-{\bf k}_F'\rangle C^{+ +}_{+ +}({\bf k}_F,{\bf k}_F',{\bf Q}),
\end{align} 
where the aforementioned extra factor of two has been accounted for.
It is illustrative to compare Eq.~(\ref{eq:less_ugly}) with its counterpart in $\delta\sigma_1$:
\begin{equation}
\label{eq:pretty}
\delta\sigma_1=-2\frac{e^2 \hbar^3}{2\pi}\frac{\tau^2}{\tau_0^2}\int\frac{dk k^2}{(2\pi)^2}\frac{k^2 v^4}{E_k^2} (G^R)^2 (G^A)^2 \frac{1}{W}\int\frac{d^2 Q}{(2\pi)^2} \sin^2\theta\cos^2\phi\, C^{+ +}_{+ +}({\bf k}_F,{\bf k}_F,{\bf Q}).
\end{equation}
In Section II we detailed the steps to follow for the evaluation of Eq.~(\ref{eq:pretty}).
Applying those same steps to Eq.~(\ref{eq:less_ugly}) and using
\begin{equation}
\int\frac{dk k^2}{2\pi^2} \frac{k^2}{E_k^2} (G^R)^2 (G^A)^2 \simeq \frac{4\pi\nu\tau_0^3}{\hbar^5 v^2}\,\,\,\mbox{ and }\,\,\,\left[\int\frac{dk k^2}{(2\pi)^2}\frac{k}{E_k} (G^A)^2 G^R\right]^2 u_0 \simeq -\frac{4\pi \nu \tau_0^3}{\hbar^5 v^2},
\end{equation}
we arrive at
\begin{equation}
\delta\sigma_2=-\frac{1}{3}\delta\sigma_1=-\frac{1}{3} G_q \ln\left(\frac{\tau_\phi}{\tau}\right).
\end{equation} 

\section{Evaluation of matrix elements for $\hat{U}$}  
\label{sec:u}

In this Appendix we calculate the coefficients entering in Eq.~(\ref{eq:coeffs}). 
These coefficients generally depend on the frequency $\Omega$ and wave vector ${\bf Q}$ of the external perturbation.
Even though only $\Omega=0$ is needed for our evaluation of $\delta\sigma$, for completeness here we allow for $\Omega\neq 0$ as well.

The calculation is facilitated by rewriting Eq.~(\ref{eq:model_b}) as
\begin{equation}
h({\bf k})=\sum_\mu\eta_\mu({\bf k}) \Lambda^\mu,
\end{equation}
where $\mu\in\{1,2,3,4\}$, $\eta_i({\bf k})= \hbar v k_i$ and $\Lambda^i=\sigma^i\tau^x$ for $i\in\{1,2,3\}$, $\eta_4({\bf k})=M$ and $\Lambda^4={\bf 1}_{2}\,\tau^z$.
Then, the finite-frequency retarded and advanced Green's functions read
\begin{equation}
\label{eq:gf}
G_{m n}^{R(A)}({\bf k},\Omega)=\frac{\epsilon^{R(A)}\delta_{m n}^0+\sum_\mu\eta_\mu \Lambda_{m n}^\mu}{[\epsilon^{R(A)}]^2-E_k^2},
\end{equation}
where $\epsilon^R\equiv\epsilon_F+i\gamma$ and $\epsilon^A\equiv\epsilon_F+\hbar\Omega-i\gamma$, with $\gamma\equiv\hbar/(2\tau_0)$ (c.f. Eq.~(\ref{eq:lifetime})).
Substituting Eq.~(\ref{eq:gf}) in Eq.~(\ref{eq:U}), we get
\begin{equation}
U^{m l}_{m' l'} = a\,\delta_{m l}\delta_{m' l'}+ \sum_\mu b_\mu\,\Lambda^\mu_{m' l'}\delta_{m l}+\sum_\mu c_\mu\,\Lambda^\mu_{m l}\delta_{m' l'}+\sum_{\mu\nu} d_{\mu\nu}\,\Lambda^\mu_{m l}\Lambda^\nu_{m' l'},
\end{equation}
where
\begin{align}
\label{eq:ints}
a&= u_0\int \frac{d^3 k}{(2\pi)^3}\frac{\epsilon^R(\epsilon^A+\hbar \Omega)}{[(\epsilon^R)^2-E_{-{\bf k}}^2][(\epsilon^A+\hbar \Omega)^2-E_{{\bf k}+{\bf Q}}^2]} \,\,\,\mbox{   ;   }\,\,\,b_\mu=u_0\int \frac{d^3 k}{(2\pi)^3}\frac{\epsilon^R d_\mu({\bf k}+{\bf Q})}{[(\epsilon^R)^2-E_{-{\bf k}}^2][(\epsilon^A+\hbar \Omega)^2-E_{{\bf k}+{\bf Q}}^2]}\nonumber\\
c_\mu&=u_0\int \frac{d^3 k}{(2\pi)^3}\frac{(\epsilon^A+\hbar \Omega) d_\mu(-{\bf k})}{[(\epsilon^R)^2-E_{-{\bf k}}^2][(\epsilon^A+\hbar \Omega)^2-E_{{\bf k}+{\bf Q}}^2]}\,\,\,\mbox{   ;   }\,\,\,d_{\mu\nu}=u_0\int \frac{d^3 k}{(2\pi)^3}\frac{d_\mu(-{\bf k}) d_\nu({\bf k}+{\bf Q})}{[(\epsilon^R)^2-E_{-{\bf k}}^2][(\epsilon^A+\hbar \Omega)^2-E_{{\bf k}+{\bf Q}}^2]},\nonumber\\
\end{align}
and  $\mu,\nu\in\{1,2,3,4\}$.
In the diffusive transport regime, namely $(\epsilon_F-M)\gg\gamma\gg (\hbar v Q,\hbar \Omega)$, the integrals in Eq.~(\ref{eq:ints}) can be analytically performed and the outcome is
\begin{align}
\label{eq:coeffs2}
a&\simeq a^{(0)}\left[1-\frac{1}{12}\left(1-\frac{M^2}{\epsilon_F^2}\right)\frac{\hbar^2 v^2 Q^2}{\gamma^2}-\frac{i\hbar \Omega}{2\gamma}\right]\nonumber\\
b_1 &=-c_1\simeq \frac{i}{6} a^{(0)}\left(1-\frac{M^2}{\epsilon_F^2}\right)\frac{\hbar v Q_x}{\gamma}\,\,\,\mbox{   ;   }\,\,\,b_2=-c_2\simeq \frac{i}{6} a^{(0)}\left(1-\frac{M^2}{\epsilon_F^2}\right)\frac{\hbar v Q_y}{\gamma}\,\,\,\mbox{   ;   }\,\,\,b_4=c_4=\frac{M}{\epsilon_F} a\nonumber\\
d_{1 1}&\simeq -\frac{1}{3}\left(1-\frac{M^2}{\epsilon_F^2}\right) a^{(0)}\left[1-\frac{1}{20}\left(1-\frac{M^2}{\epsilon_F^2}\right)\frac{(3 Q_x^2+Q_y^2) \hbar^2 v^2}{\gamma^2}-\frac{i\hbar \Omega}{2\gamma}\right]\nonumber\\
d_{2 2}&\simeq -\frac{1}{3}\left(1-\frac{M^2}{\epsilon_F^2}\right) a^{(0)}\left[1-\frac{1}{20}\left(1-\frac{M^2}{\epsilon_F^2}\right)\frac{(3 Q_y^2+Q_x^2) \hbar^2 v^2}{\gamma^2}-\frac{i\hbar \Omega}{2\gamma}\right]\nonumber\\
d_{3 3}&\simeq -\frac{1}{3}\left(1-\frac{M^2}{\epsilon_F^2}\right) a^{(0)}\left[1-\frac{1}{20}\left(1-\frac{M^2}{\epsilon_F^2}\right)\frac{\hbar^2 v^2 Q^2}{\gamma^2}-\frac{i\hbar \Omega}{2\gamma}\right]\nonumber\\
d_{4 4}&\simeq\frac{M^2}{\epsilon_F^2} a\nonumber\\
d_{1 2} &=d_{2 1}\simeq a^{(0)}\frac{1}{30}\left(1-\frac{M^2}{\epsilon_F^2}\right)^2\frac{\hbar^2 v^2 Q_x Q_y}{\gamma^2}\nonumber\\
d_{1 4}&=-d_{4 1}\simeq -i a^{(0)}\frac{M}{\epsilon_F}\left(1-\frac{M^2}{\epsilon_F^2}\right)\frac{\hbar v Q_x}{\gamma}
\,\,\,\mbox{   ;   }\,\,\,d_{2 4}=-d_{4 2}\simeq -i a^{(0)}\frac{M}{\epsilon_F}\left(1-\frac{M^2}{\epsilon_F^2}\right)\frac{\hbar v Q_y}{\gamma},
\end{align}
where $a^{(0)}\equiv[2 (1+M^2/\epsilon_F^2)]^{-1}$, and the elements omitted above are zero.
It is worth noting that Eq.~(\ref{eq:coeffs2}) can be used to investigate the dynamical spin-charge coupling in doped TIs.
Since this task is not directly related to the theme of this paper, it will be pursued elsewhere. 

\section{Classical conductivity of two  coupled layers}
\label{sec:cond}

In this Appendix we analyze the classical conductivity of two coupled layers.
The current in layer $i$ is given by ${\bf j}_i=\sum_j \sigma_{i j} {\bf E}_j$.
It is illustrative to write $\sigma_{i j}$ in terms of the diffusive density-density response, using the continuity equation
\begin{equation}
\frac{\partial\rho_i}{\partial t}+\nabla\cdot{\bf j}+\lambda\sum_j (\rho_j-\rho_i)=0
\end{equation}
along with the constitutive equation ${\bf j}_i=-D_i {\boldsymbol\nabla}\rho_i-e^2\nu_i D_i {\bf E}_i$.
$\lambda$ is the interlayer tunneling rate. 
Thus it follows that
\begin{equation}
\sigma_{i j}({\bf q},\omega)=-\frac{i\omega}{q^2} \chi_{i j}+\frac{\lambda}{q^2}\sum_k (\chi_{i j}-\chi_{k j}),
\end{equation}
where $\chi_{i j} ({\bf q},\omega)=e^2 \nu_j D_j q^2 p_{i j}({\bf q},\omega)$ is the density-density response function and 
\begin{equation}
p_{i j}({\bf q},\omega) = \left\{\begin{array}{ccc} \tilde{p}_i^{(0)}/(1-\lambda^2 p_1^{(0)} p_2^{(0)}) & {\rm if } & i=j\\ 
                                   \lambda \tilde{p}_1^{(0)} \tilde{p}_2^{(0)}/(1-\lambda^2 p_1^{(0)} p_2^{(0)}) & {\rm if } & i\neq j\end{array}\right.,
\end{equation}
with $\tilde{p}_i^{(0)} \equiv(D_i q^2-i\omega+\lambda)^{-1}$. 
The dressed diffusion probability $p_{i i}$, derived here from the continuity equation, has identical form as Eq.~(\ref{eq:cii}), which was derived microscopically in Section IIIB.
Here $\omega$ and ${\bf q}$ are the frequency and momentum associated with the applied electric field. 
A straightforward calculation shows that $\sigma_{1 2}=\sigma_{2 1}=0$ when ${\bf E}_i$ is spatially uniform (${\bf q}=0$).

\section{Equations for coupled Cooperons}
\label{sec:coupled_cooper}

In the first part of this Appendix we present an alternative derivation for the results of Section IIIB. 
In the second part of the Appendix we generalize the derivation to make it suitable for TI thin films with $\tau_{\phi 1}\ll\tau_v$, which contain two gapless singlet Cooperons in the bulk and one gapless singlet Cooperon on the surface.
The outcome of such generalization is the third line of Eq.~(\ref{eq:res_tot}).

\subsection{Two 2D layers without spin-orbit coupling}

In this subsection we use ``1'' and ``2'' to label the two layers.
The relevant Cooperon modes are then $C_{1 1}$, $C_{1 2}$, $C_{2 1}$ and $C_{2 2}$.
Recognizing that Cooperons must obey a diffusion equation in absence of phase relaxation, we posit the following coupled equations:
\begin{align}
\label{eq:1}
(D_1 Q^2 +\tau_{\phi 1}^{-1}) C_{1 1}+\lambda (C_{1 1}-C_{2 1}) &=\hbar/(2\pi\nu_1\tau_{d 1}^2)\nonumber\\
(D_2 Q^2+\tau_{\phi 2}^{-1}) C_{2 1}+\lambda (C_{2 1}-C_{1 1}) &= 0\nonumber\\
(D_2 Q^2+\tau_{\phi 2}^{-1}) C_{2 2}+\lambda(C_{2 2}-C_{1 2}) &=\hbar/(2\pi\nu_2\tau_{d 2}^2)\nonumber\\
(D_1 Q^2+\tau_{\phi 1}^{-1}) C_{1 2}+\lambda (C_{1 2}-C_{2 2})&=0,
\end{align}
where $\lambda$ is the interlayer tunneling rate.
Note that the source term appears only for the diagonal terms of the $2\times 2$ Cooperon matrix.
The solution of Eq.~(\ref{eq:1}) reads
\begin{align}
C_{1 1}&=\frac{\hbar}{2\pi\nu_1\tau_1^2}\frac{D_2 Q^2+\tilde{\tau}_{\phi 2}^{-1}}{(D_1 Q^2+\tilde{\tau}_{\phi 1}^{-1})(D_2 Q^2+\tilde{\tau}_{\phi 2}^{-1})-\lambda^2}\nonumber\\
C_{2 2}&=\frac{\hbar}{2\pi\nu_2\tau_2^2}\frac{D_1 Q^2+\tilde{\tau}_{\phi 1}^{-1}}{(D_1 Q^2+\tilde{\tau}_{\phi 1}^{-1})(D_2 Q^2+\tilde{\tau}_{\phi 2}^{-1})-\lambda^2}\nonumber\\
C_{1 2}&=C_{2 1}=\frac{\lambda}{D_2 Q^2+\tilde{\tau}_{\phi 2}^{-1}} C_{1 1},
\end{align}
where $\tilde{\tau}_{\phi i}^{-1}\equiv \tau_{\phi i}^{-1}+\lambda$.
The expressions for $C_{1 1}$ and $C_{2 2}$ agree with Eq.~(\ref{eq:cii}).
In addition, $C_{1 2}$ and $C_{2 1}$ agree with the expressions for $p_{1 2}$ and $p_{2 1}$ derived in Appendix~\ref{sec:cond} (where we discussed the classical diffusive conductivity).
$C_{i i}$ of Eq.~(\ref{eq:1}) is equivalent to $C^{i i}_{i i}$ of Fig.~\ref{fig:dsij}. 
Likewise, $C_{1 2}$ and $C_{2 1}$ of Eq.~(\ref{eq:1}) correspond to $C^{1 1}_{2 2}$ and $C^{2 2}_{1 1}$ of Fig.~\ref{fig:dsij}.
Although $C_{1 2}$ and $C_{2 1}$ are nonzero, they do not contribute to $\delta\sigma$ because the velocity operator is diagonal in the layer index. 
Therefore, we reproduce the expression of Section IIIB for $\delta\sigma$.

\subsection{TI film with two gapless bulk Cooperons and one gapless surface Cooperon}

In this subsection we use ``1'' and ``3'' to label the two bulk Cooperons, and ``2'' to label the surface Cooperon.
The generalization of Eq.~(\ref{eq:1}) is
\begin{align}
\label{eq:2}
(D_1 Q^2+\tau_{\phi 1}^{-1}) C_{1 1}+\lambda(C_{1 1}-C_{2 1})&=\hbar/(2\pi\nu_1\tau_{d 1}^2)\nonumber\\
(D_2 Q^2+\tau_{\phi 2}^{-1}) C_{2 1}+\lambda(2 C_{2 1}-C_{1 1}-C_{3 1})&=0\nonumber\\
(D_1 Q^2+\tau_{\phi 1}^{-1}) C_{3 1}+\lambda(C_{3 1}-C_{2 1})&=0,
\end{align}
\begin{align}
\label{eq:3}
(D_1 Q^2+\tau_{\phi 1}^{-1}) C_{1 2}+\lambda (C_{1 2}-C_{2 2}) &= 0\nonumber\\
(D_2 Q^2+\tau_{\phi 2}^{-1}) C_{2 2}+\lambda( 2 C_{2 2}-C_{1 2}-C_{3 2})&=\hbar/(2\pi\nu_2\tau_{d 2}^2)\nonumber\\
(D_1 Q^2+\tau_{\phi 1}^{-1}) C_{3 2}+\lambda(C_{3 2}-C_{2 2})&=0
\end{align}
and 
\begin{align}
\label{eq:4}
(D_1 Q^2+\tau_{\phi 1}^{-1}) C_{1 3}+\lambda(C_{1 3}-C_{2 3})&=0\nonumber\\
(D_2 Q^2+\tau_{\phi 2}^{-1}) C_{2 3}+\lambda(2 C_{2 3}-C_{1 3}-C_{3 3})&=0\nonumber\\
(D_1 Q^2+\tau_{\phi 1}^{-1}) C_{3 3}+\lambda(C_{3 3}-C_{2 3})&=\hbar/(2\pi\nu_1\tau_{d 1}^2),
\end{align}
Once again in Eqs.~(\ref{eq:2})-(\ref{eq:4}) the source term appears for the diagonal components of the $3\times 3$ Cooperon matrix.
In addition, a factor of $2$ has been multiplied in front of some tunneling rates associated to surface Cooperons. 
The rationale behind this is that the Cooperon on the surface can decay into two bulk modes, i.e. the effective decay rate becomes $\tau_{\phi 2}^{-1}+2\lambda$.
Aside from this, we have assumed a unique tunneling rate $\lambda$ between all pairs of Cooperons.

The quantum correction to conductance can be written as
\begin{equation}
\label{eq:ds_app}
\delta G=2\frac{e^2}{\hbar^2}\nu_1 D_1 \tau_{d 1}^2\int_{\bf Q} (C_{1 1}+C_{3 3})+ 2\frac{e^2}{\hbar^2}\nu_2 D_2 \tau_{d 2}^2\int_{\bf Q} C_{2 2}.
\end{equation}

Solving Eqs.~(\ref{eq:2})-(\ref{eq:4}) requires some algebra.
The results for the Cooperons of interest are
\begin{align}
\label{eq:cii_app}
C_{1 1}&=C_{3 3}=\frac{\hbar}{2\pi\nu_1\tau_{d 1}^2}\frac{(D_1 Q^2+\tilde{\tau}_{\phi 1}^{-1})(D_2 Q^2+\tilde{\tau}_{\phi 2}^{-1}+\lambda)-\lambda^2}{(D_1 Q^2+\tilde{\tau}_{\phi 1}^{-1})\left[(D_1 Q^2+\tilde{\tau}_{\phi 1}^{-1})(D_2 Q^2+\tilde{\tau}_{\phi 2}^{-1}+\lambda)-2 \lambda^2\right]}\nonumber\\
C_{2 2} &=\frac{\hbar}{2\pi\nu_2\tau_{d 2}^2}\frac{D_1 Q^2+\tilde{\tau}_{\phi 1}^{-1}}{(D_1 Q^2+\tilde{\tau}_{\phi 1}^{-1})(D_2 Q^2+\tilde{\tau}_{\phi 2}^{-1}+\lambda)-2 \lambda^2},
\end{align}
which are not illuminating expressions.
It is better to rewrite them as
\begin{align}
\label{eq:ciib_app}
C_{1 1} &=C_{3 3}=\frac{\hbar}{2\pi\nu_1\tau_{d 1}^2}\frac{1}{D_1}\left[\frac{X}{Q^2+q_x^2}+\frac{Y}{Q^2+q_y^2}+\frac{Z}{Q^2+q_z^2}\right]\nonumber\\
C_{2 2} &=\frac{\hbar}{2\pi\nu_2\tau_{d 2}^2} \frac{1}{D_2}\left[\frac{A}{Q^2+q_a^2}+\frac{B}{Q^2+q_b}\right],
\end{align}
so that Eq.~(\ref{eq:ds_app}) transforms into
\begin{equation}
\label{eq:ds2_app}
\delta G=\frac{e^2}{\pi\hbar}\int_{\bf Q} \left[2\frac{X}{Q^2+q_x^2}+2\frac{Y}{Q^2+q_y^2}+2\frac{Z}{Q^2+q_z^2}+\frac{A}{Q^2+q_a^2}+\frac{B}{Q^2+q_b^2}\right].
\end{equation}

Comparing Eqs.~(\ref{eq:cii_app}) and ~(\ref{eq:ciib_app}), we arrive at
\begin{align}
\label{eq:A_app}
A &=\frac{\frac{1}{D_1\tilde{\tau}_{\phi 1}}-q_a^2}{q_b^2-q_a^2}\,\,\,\mbox{   ;   }\,\,\,B = 1-A\nonumber\\
X &= \frac{(D_1 q_x^2-\tilde{\tau}_{\phi 1}^{-1})(D_2 q_x^2-\tilde{\tau}_{\phi 2}^{-1}-\lambda)-\lambda^2}{D_1 D_2 (q_x^2-q_y^2)(q_x^2-q_z^2)}\nonumber\\
Y &= \frac{ D_2 q_y^2 \tilde{\tau}_{\phi 1}^{-1}-\tilde{\tau}_{\phi 1}^{-1}(\tilde{\tau}_{\phi 2}^{-1}+\lambda)+D_1 q_y^2 (-D_2 q_y^2+\tilde{\tau}_{\phi 2}^{-1}+\lambda)+\lambda^2}{D_1 D_2 (q_x^2-q_y^2)(q_y^2-q_z^2)}\nonumber\\
Z &= \frac{ D_2 q_z^2 \tilde{\tau}_{\phi 1}^{-1}-\tilde{\tau}_{\phi 1}^{-1}(\tilde{\tau}_{\phi 2}^{-1}+\lambda)+D_1 q_z^2 (-D_2 q_z^2+\tilde{\tau}_{\phi 2}^{-1}+\lambda)+\lambda^2}{D_1 D_2 (q_x^2-q_z^2)(q_z^2-q_y^2)}
\end{align}
and
\begin{align}
\label{eq:qa_app}
2 q_{a(b)}^2 &=\frac{1}{D_1\tilde{\tau}_{\phi 1}}+\frac{1}{D_2\tilde{\tau}_{\phi 2}}+\frac{\lambda}{D_2}\pm\sqrt{\left(\frac{1}{D_1\tilde{\tau}_{\phi 1}}-\frac{1}{D_2 \tilde{\tau}_{\phi 2}}-\frac{\lambda}{D_2}\right)^2+\frac{8\lambda^2}{D_1 D_2}}\nonumber\\
q_{x (y)}^2 &=q_{a (b)}^2\,\,\,\mbox{   ;   }\,\,\,q_z^2 = 1/(D_1\tilde{\tau}_{\phi 1}).
\end{align}
Note that $q_{a (b)}=q_{x (y)}$, which will be important below.
Also note that the expressions for $A$, $B$ and $q_{a(b)}$ are identical to the ones in Section IIIB, except for the following difference: the effective inelastic scattering rate for layer $2$ is now $\tau_{\phi 2}^{-1}+2\lambda$ instead of $\tau_{\phi 2}^{-1}+\lambda$, for the reason explained above.

Although Eqs. ~(\ref{eq:A_app}) and ~(\ref{eq:qa_app}) look cumbersome, after substituting Eq.~(\ref{eq:qa_app}) back in Eq.~(\ref{eq:A_app}) we find some remarkable simplifications.
In particular
\begin{equation}
Z=1/2\,\,\,\mbox{  ,   }\,\,\,2 X+ A=1 \,\,\,\mbox{    and   }\,\,\, 2 Y+ B =1.
\end{equation}
Replacing these in Eq.~(\ref{eq:ds2_app}) immediately leads to
\begin{equation}
\delta G=\frac{e^2}{\pi\hbar}\int_{\bf Q} \left[\frac{1}{Q^2+q_a^2}+\frac{1}{Q^2+q_b^2}+\frac{1}{Q^2+q_z^2}\right].
\end{equation}
In consequence, we recover the third line of Eq.~(\ref{eq:res_tot}) for the low-field magnetoconductance:
\begin{equation}
\label{eq:dg}
\frac{\Delta G}{G_q}=\frac{1}{2}\left[f\left(\frac{H_a}{H}\right)+f\left(\frac{H_b}{H}\right)+f\left(\frac{H_z}{H}\right)\right],
\end{equation}
where $H_a=\hbar q_a^2/(4 e)$, etc. 
As a reality check, let us take some simple limits.

First, consider the case of no bulk-surface coupling, $\lambda\to 0$.
In this case $H_a=H_z=\hbar/(4 e D_1 \tau_{\phi 1})$ and $H_b=\hbar/(4 e D_2 \tau_{\phi 2})$, which produces 
\begin{equation}
\frac{\Delta G}{G_q}=\frac{1}{2}\left[2 f\left(\frac{H_a}{H}\right)+f\left(\frac{H_b}{H}\right)\right].
\end{equation}
This is indeed the result one would have expected when bulk and surface are decoupled.

Second, suppose both $\tau_{\phi 1}$ and $\tau_{\phi 2}$ are infinitey large, for arbitrary tunneling rate.
Then it follows that $H_b=0$,
\begin{equation}
H_a=\frac{\hbar}{4 e}\lambda\left(\frac{1}{D_1}+\frac{2}{D_2}\right)\,\,\,\mbox{   and   }\,\,\, H_z=\frac{\hbar}{4 e}\frac{\lambda}{D_1}
\end{equation}
Then, 
\begin{equation}
\frac{\Delta G}{G_q}=\frac{1}{2}\left[f\left(\frac{H_a}{H}\right)+f\left(\frac{H_z}{H}\right)\right].
\end{equation}
The fact that $H_b=0$ means that we recover the conventional WAL case (as we should when the phase relaxation times are infinitely long).

Finally, consider the case of very strong tunneling between bulk and surface states.
In this case $H_a$ and $H_z$ become very large ($\propto\lambda$), whereas $H_b$ becomes independent of $\lambda$. 
Consequently
\begin{equation}
\frac{\Delta G}{G_q}=\frac{1}{2} f\left(\frac{H_b}{H}\right),
\end{equation}
as if we had a single channel contributing to WAL. 
This seems to make sense too, because when tunneling is strong, $C_{i i}$ are strongly coupled to one another ($i=1,2,3$).

\section{Some special cases of Eq.~(\ref{eq:res_tot})}
\label{sec:special}

In this Appendix we analyze some simple limiting cases of Eq.~(\ref{eq:res_tot}), which considers a single TI surface coupled to bulk states.
First, suppose that surface-bulk tunneling is strong, so that $\tau_{t i}\ll \tau_{\phi i}$ for $i=1,2$.
In this case $(H_a, H_c, \tilde{H}_1)\gg (H_b,H_d)$ and thus Eq.~(\ref{eq:res_tot}) turns into
\begin{align}
\label{eq:res_tot2}
& \frac{\Delta G}{G_q}=\frac{1}{2}\left\{\begin{array}{ccc} 
f(H_b/H) &{\rm if } & \tilde{\tau}_H\ll\tau_s\\
f(H_b/H) &{\rm if } & \tilde{\tau}_H\gg(\tau_v,\tau_s)\\
f(H_d/H) &{\rm if } & \tilde{\tau}_H\ll\tau_v,
\end{array}\right.
\end{align}
where $H_b \simeq \hbar/(4 e)(1/\tau_{\phi 1}+1/\tau_{\phi 2})/(D_1+D_2)$ and $H_d\simeq \hbar/(4 e)(2/\tau_{\phi 1}+1/\tau_{\phi 2})/(2 D_1+D_2)$. 
For simplicity we have taken $\tau_{t 1}=\tau_{t 2}$, but this assumption can be easily relaxed.
In sum, WL is {\em not} possible when the bulk-surface coupling is strong, and the film exhibits conventional WAL ($\alpha=1/2$) regardless of the bulk carrier concentration. 

Next, we consider a weak surface-bulk tunneling, so that $\tau_{t i}\gg \tau_{\phi i}$ for $i=1,2$.
In this case the outcome depends on whether $D_1\tau_{\phi 1}>D_2\tau_{\phi 2}$ or $D_1\tau_{\phi 1}<D_2\tau_{\phi 2}$.
Without loss of generality suppose that $D_1\tau_{\phi 1}>D_2\tau_{\phi 2}$.
Then Eq.~(\ref{eq:res_tot}) yields
\begin{align}
\label{eq:res_tot3}
& \frac{\Delta G}{G_q}\simeq\frac{1}{2}\left\{\begin{array}{ccc} 
f(H_{\phi 2}/H)-2 f(H_{\phi 1}/H) &{\rm if } & \tilde{\tau}_H\ll\tau_s\\
f(H_{\phi 2}/H)+f(H_{\phi 1}/H) &{\rm if } & \tilde{\tau}_H\gg(\tau_v,\tau_s)\\
f(H_{\phi 2}/H)+2 f(H_{\phi 1}/H) &{\rm if } & \tilde{\tau}_H\ll\tau_v,
\end{array}\right.
\end{align}
where $H_{\phi i}= \hbar/(4 e D_i \tau_{\phi i})$ for $i=1,2$.
When $H_{\phi 1}$ and $H_{\phi 2}$ are of the same order, the first line of Eq.~(\ref{eq:res_tot3}) displays WL with $\alpha=-1/2$ and the third line exhibits WAL with $\alpha=3/2$. 
If instead $H_{\phi 1}\ll H_{\phi 2}$, $\Delta G$ is the same as if there were no surface states.
This latter regime can be experimentally accessible by e.g. depositing magnetic impurities on the surface of the TI.

Last, we consider the case $\tau_{t 1}\gg\tau_{\phi i}\gg\tau_{t 2}$ for $i=1,2$.
This situation may be relevant for some thicker TI films where $\tau_{t 1}/\tau_{t 2}=W\nu_1/\nu_2\gg 1$ (for thicker films, surface states have more bulk states to decay onto).
The resulting magnetoconductance is once again as though there were no surface states: 
\begin{align}
\label{eq:res_tot5bis}
& \frac{\Delta G}{G_q}=\left\{\begin{array}{ccc} 
-f(H_{\phi 1}/H) &{\rm if } & \tilde{\tau}_H\ll\tau_s\\
\frac{1}{2}f(H_{\phi 1}/H) &{\rm if } & \tilde{\tau}_H\gg(\tau_v,\tau_s)\\
f(H_{\phi 1}/H) &{\rm if } & \tilde{\tau}_H\ll\tau_v.
\end{array}\right.
\end{align}
\end{widetext}

\end{document}